\documentclass{aa}  

\usepackage{graphicx}
\usepackage{caption}
\usepackage{txfonts}
\usepackage{float}
\usepackage[hyperindex, breaklinks, colorlinks, linkcolor=blue, citecolor=blue]{hyperref}

\usepackage{capt-of}
\usepackage{placeins}
\usepackage{dblfloatfix}
\usepackage{xcolor}
\usepackage{color}
\usepackage{afterpage}
\usepackage{orcidlink}
\usepackage{cuted}
\usepackage{needspace}
\newcommand{\citeyearless}[1]{\citeauthor{#1} \citeyear{#1}}
\titlerunning{A three-step approach to reliably estimate magnetic field strengths in star-forming regions}
\authorrunning{A. Polychronakis et al.}

\usepackage{subcaption}

\begin{document} 

    \title{A three-step approach to reliably estimate magnetic field strengths in star-forming regions}

   % \title{Estimate the magnetic field strength in star-forming regions in three easy steps}

   % \subtitle{}

\author{
  A.~Polychronakis\inst{1,2}\orcidlink{0009-0005-7962-6296}
  \and
  A.~Tritsis\inst{3}\orcidlink{0000-0003-4987-7754}
  \and
  R.~Skalidis\inst{4}\thanks{Hubble fellow}\orcidlink{0000-0003-2337-0277}
  \and
  K.~Tassis\inst{1,2}\orcidlink{0000-0002-8831-2038}
}

    \institute{Department of Physics, University of Crete, GR-70013, Heraklion, Greece\\
          \email{ph5698@edu.physics.uoc.gr} % Email just under the first university
          \and Institute of Astrophysics, Foundation for Research and Technology-Hellas, Vasilika Vouton, GR-70013 Heraklion, Greece
          \and Institute of Physics, Laboratory of Astrophysics, Ecole Polytechnique Fédérale de Lausanne (EPFL), Observatoire de Sauverny, 1290, Versoix, Switzerland
          \and TAPIR, California Institute of Technology, MC 350-17, Pasadena, CA 91125, USA
          }

    \date{Received date; accepted date}

% \abstract{}{}{}{}{} 
% 5 {} token are mandatory
 
  \abstract
  % {} % leave it empty if necessary  
   {The magnetic field has been shown to play a crucial role in star formation. Dust polarization is one of the most effective tools for probing the properties of the magnetic field, yet it does not directly trace its strength. To bridge this gap, several methods have been developed, combining polarization and spectroscopic data, to estimate the strength of the magnetic field. The most widely applied method was developed by \cite{DAVIS} and \cite{FERMI}, hereafter DCF, and relates the polarization angle dispersion to magnetic field strength under the assumption of Alfvénic turbulence. \cite{ST}, hereafter ST, relaxed this assumption to account for the compressible modes, deriving more accurate estimates of the magnetic field strength than the DCF in clouds with no self-gravity. The accuracy of these methods in self-gravitating regions is poorly explored.}
   {We aim to evaluate the accuracy of these magnetic-field estimation methods in star-forming regions and propose a systematic approach for calculating the key observational parameters they involve: the velocity dispersion ($\delta$v), the polarization angle dispersion ($\delta\theta$), and the cloud density ($\rho$).}
   {We use a 3-dimensional magnetohydrodynamic chemo-dynamical simulation of a turbulent collapsing molecular cloud. We generate synthetic observations, for seven different inclination angles with respect to the mean component of the magnetic field, that encompass a comprehensive set of observables, including emission line spectra, Stokes parameters, and column density maps. We employ various approaches for estimating the parameters $\delta$v, $\delta \theta$, and $\rho$, and identify the best approach that most effectively probes the plane-of-sky (POS) component of the magnetic field.}
   {We find that the approach used to calculate the parameters $\delta$v, $\delta \theta$, and $\rho$ plays a crucial role in estimating the magnetic field strength, regardless of the specific method used (i.e., the DCF or the ST methods). We show that the value probed by both methods corresponds to the median of the molecular-species--weighted POS component of the magnetic field. We also find that ST outperforms DCF. The magnetic field strength values derived with the ST method accurately follow the expected cosine trend with respect to the inclination angle of the magnetic field, and consistently remain within 1$\sigma$ of the median component of the magnetic field strength. In self-gravitating clouds, we propose the following approach to accurately constrain the intrinsic parameters involved in the magnetic field estimation methods: $\rho$ using radiative transfer analysis, $\delta$v using the second moment maps, and $\delta \theta$ by fitting Gaussians to the polarization angle distributions to remove the contribution of the hourglass morphology.}
   {}


   \maketitle

%
%-------------------------------------------------------------------

\section{Introduction}

The magnetic field of the interstellar medium (ISM) plays a crucial role in regulating star formation, influencing the dynamics of gas, and shaping the structure of the ISM (e.g., \citealt{CRUTCHER, TRITSIS2015, HENNEBELLE, MAURY, TSUKAMOTO, PATTLE2023}). One of the most effective ways to probe the magnetic field in these clouds is through observations of dust polarization, which traces the orientation of the plane-of-sky (POS) magnetic field (e.g., {\citealt{ANDERSSON}}). With the use of polarization data, we can estimate the strength of the magnetic field, which is a key parameter to asses its dynamical importance in comparison to other forces exerted in the ISM.

\cite{DAVIS} and \cite{FERMI}, hereafter DCF, were the first to propose a method to estimate the magnetic field strength from dust polarization observations. This approach relates the dispersion of the polarization angles to the magnetic field strength under the assumption of Alfvénic turbulence. Since then, several corrections and modifications have been proposed to the DCF method (e.g., \citealt{ZWEIBEL, MYERS, OSTRIKER, PADOAN, HEITSCH, KUDOH, GIRART2006, FALCETA, HILDEBRAND, HOUDE, YOO, PATTLE}).

A more recent approach developed by \cite{ST}, hereafter ST, accounted for the the presence of compressible magnetohydrodynamic (MHD) modes, which have been neglected by DCF. This new method provides more accurate estimates of the magnetic field strength in a wider range of astrophysical environments (\citealt{SKALIDIS}). Before diving into the specifics of each method, it is essential to understand their foundational principles and the contexts in which they are applied.

DCF posited that magnetic field lines are distorted by the propagation of incompressible transverse MHD waves, known as Alfv\'en waves. This distortion causes a spread in the polarization angle distribution, which, when combined with measurements of the gas turbulent motions from spectroscopic data, allows for the estimation of the magnetic field strength.

The magnetic field can be decomposed into a mean and a fluctuating component, $\mathbf{B_0}$ and $\mathbf{\delta B}$ respectively. Therefore, the total field is $\mathbf{B = B_0 + \delta B}$, and the total magnetic energy density is
\begin{equation}\label{eq1}
\rm{\frac{B^2}{8\pi} = \frac{1}{8\pi}[B_0^2 + \delta B^2 + 2\mathbf{\delta B}\cdot \mathbf{B_0}]}.
\end{equation}

The DCF method assumes infinite conductivity in the ISM, implying that the magnetic field is ``frozen-in'' the gas, such that the gas and the magnetic field oscillate together. Turbulent gas motions induce small fluctuations in the magnetic field, \( |\delta \mathbf{B}| \ll |\mathbf{B_0}| \), treated as Alfv\'en waves, and since Alfv\'en waves are transverse, we have \(\mathbf{B_0} \cdot \delta \mathbf{B} = 0\). The method assumes that the kinetic energy of these turbulent motions equals the energy density of the magnetic field fluctuations

\begin{equation}
\label{eq:dcf_equipartition}
\rm{\frac{1}{2}\rho\delta\text{v}^2 = \frac{\delta B^2}{8\pi}},
\end{equation}
where \(\rho\) is the gas density and \(\delta v\) the rms velocity. Dividing both sides by \(\rm{B_0^2}\) and rearranging, we obtain
\begin{equation}
\rm{B_0 = \sqrt{4\pi\rho} \, \delta\text{v} \left[\frac{\delta B}{B_0}\right]^{-1}}.
\end{equation}

Dust polarization traces the magnetic field orientation, with a \(\pi\) ambiguity, and the spread of the polarization angle \(\delta \theta\) reflects the ratio \(\rm{\delta B / B_0}\). A strong mean field keeps \(\delta \theta\) small, as the field lines remain weakly perturbed, while larger fluctuations lead to an increase in \(\delta \theta\) due to turbulence-induced dispersion. Considering that to first order \(\rm{\delta \theta \approx \delta B / B_0}\), we get

\begin{equation}
\rm{B_0 = \sqrt{\frac{4\pi\rho}{3}}\frac{\delta\text{v}}{\delta\theta}},
\end{equation}
where $1/\sqrt{3}$ was inserted by DCF under the assumption that there is only one velocity component associated with the dispersion of the field lines and that the turbulent motions are isotropic. In the general case, we can write the DCF method as
\begin{equation}
\rm{{B_0 = f\sqrt{4\pi\rho}\frac{\delta\text{v}}{\delta\theta}}},
\label{DCF_eq}
\end{equation}
where f is a correction factor. Based on MHD simulations, different correction factors f have been proposed which are usually somewhat different than 1/$\sqrt{3}$ (e.g., \citealt{OSTRIKER}) and also account for integration over sub-beam regions and line-of-sight (LOS) structures (see \citealt{LIU2021}).

Motivated by the lack of inclusion of compressible modes in the DCF method, {\cite{ST}} proposed a more general method that accounts for these modes. Starting from Eq.{\ref{eq1}} and assuming, like DCF, that the gas is perfectly coupled to the magnetic field and turbulent motions fully transfer to magnetic fluctuations, they differentiated from the DCF by including all MHD modes. ST considered the cross-term in magnetic energy, \(\rm{\delta B \cdot B_0}\). When \(\rm{|\delta B| \ll |B_0|}\), to leading order, the fluctuating part in the magnetic energy equation will be dominated by the root mean square of $\rm{\delta B \cdot B_0}$, as shown by numerical (e.g., {\citealt{federrath2016, BEATTIE, BEATTIE2022}) and analytical calculations (\citealt{SKALIDIS2023B}). Thus, the fluctuating part in the energy equation, will be

\begin{equation}
\rm{\delta\epsilon_{m} \approx \frac{\sqrt{\langle\delta \rm{B_{\parallel}\rangle^2}} \, B_0}{4\pi}}
,
\end{equation}
where $\sqrt{\langle\delta \rm{B_{\parallel}\rangle^2}}$ is the root mean square magnetic field perturbation parallel to the mean magnetic field $\rm{\mathbf{B_0}}$.

Assuming that the turbulent kinetic energy is equal to the magnetic energy fluctuations, they derived

\begin{equation}
\label{eq:coupling_term}
\rm{\frac{1}{2}\rho\delta\text{v}^2 \approx \frac{\sqrt{\langle\delta \rm{B_{\parallel}\rangle^2}} B_0}{4\pi}}.
\end{equation}
Finally, replacing $\delta\theta = \rm{\sqrt{\langle\delta \rm{B_{\parallel}\rangle^2}}/B_0}$ as in DCF, they obtained
\begin{equation}
\rm{B_0 = \sqrt{2\pi\rho}\frac{\delta\text{v}}{\sqrt{\delta\theta}}}.
\label{ST_eq}
\end{equation}

The DCF and ST methods differ in several key aspects. First, the ST method scales with the square root of $\delta\theta$, which affects the resulting magnetic field strength estimates. The DCF method includes a correction factor, f, which is not consistently set across all observational studies, potentially leading to biases in statistical studies. In contrast, ST method was found to be consistent with a number of numerical simulations with different initial conditions (hence a different number of turbulent eddies along the LOS) and different physical resolutions, without a need for correction factor. Additionally, the ST method has been shown to perform well in turbulent MHD simulations (e.g., {\citealt{SKALIDIS}}), in the absence of self-gravity, highlighting its robustness in environments dominated by turbulence. However, both the ST and DCF methods, by their design, omit the effects of self-gravity. The gravitational collapse will contribute to both the velocity dispersion and the dispersion of polarization angles. The latter is true due to the expected hourglass-like shape of the magnetic field lines which has been observed in many clouds and cores (e.g., \citealt{GIRART2006, GIRART2009, FRAU2011, STEPHENS, QIU, BELTRAN, SAHA}). Therefore, such effects should also be considered when applying these methods to self-gravitating regions. However, there is currently no standardized or well-established approach for incorporating this correction.

Numerous applications of the methods have been carried out on diffuse clouds (e.g., {\citealt{PANOPOULOU, SKALIDIS2022, KALBERLA, SKALIDIS2023}}) as well as in self-gravitating regions (e.g., \citealt{LIU2019, COUDE, PALAU2021, PATTLE2021, ARZOUMANIAN2021, ESWARAIAH, GONG, KWON, HWANG, WARD, KAROLY, JONATHAN, HU, CHOI}}), establishing a robust framework in the literature. In all the aforementioned studies, observers employ a variety of methods to estimate the three key parameters \(\delta\theta\), \(\delta\)v, and \(\rho\). This methodological ambiguity can introduce uncertainties that affect the obtained estimates of the magnetic field strength. Recent advancements, such as the machine learning approach introduced by \cite{JONATHAN} have sought to improve magnetic field strength estimation in self-gravitating regions. While their method demonstrates high precision by leveraging synthetic data and advanced algorithms, it requires intricate models and extensive training. In contrast, the three-step approach for calculating (\(\delta\theta\), \(\delta\)v, and \(\rho\)) proposed in this paper, offers a simpler, yet still accurate, methodology for calculating the POS component of the magnetic field in these regions.

\begin{figure*}[t] % 't' for top placement
    \centering
    \includegraphics[width=1.\textwidth,height=0.9\textheight,keepaspectratio]{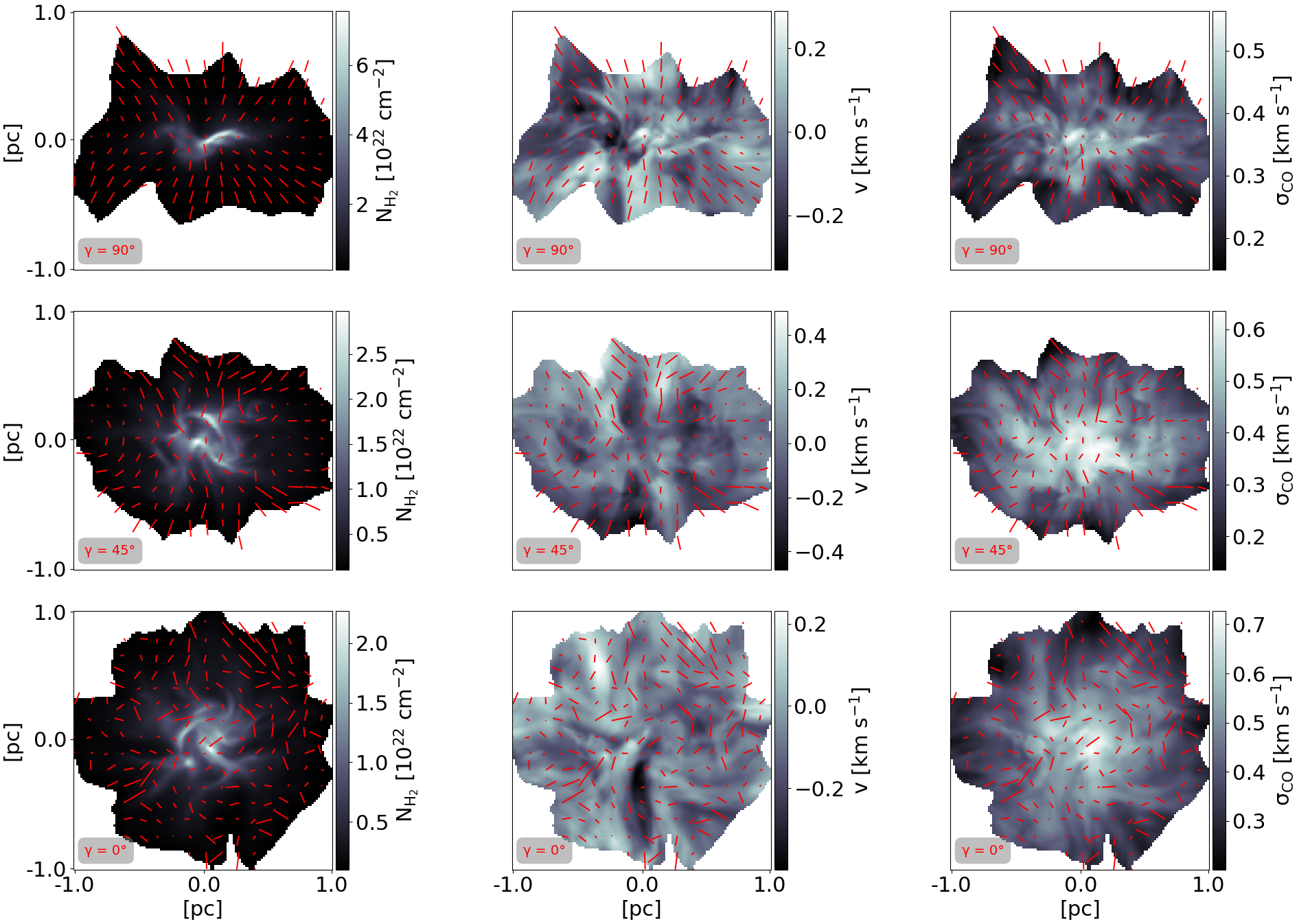} 
    \caption{Column density (left column), first moment (middle column) and second moment maps (right column) for three inclination angles ($\gamma$). Red lines correspond to the polarization pseudo-vectors. To generate the moment maps we use the CO $\rm{J = 1\rightarrow0}$ line. The modeled cloud was analyzed at a time corresponding to 1.2 times the free fall time, at which stage the central number density has reached a value of \( 10^5 \, \text{cm}^{-3} \).}
    \label{maps}
\end{figure*} 

A key novelty of our study lies in the data used. We employ mock polarization and true non-local thermodynamic equilibrium (non-LTE) radiative transfer position-position-velocity (PPV) cubes based on chemo-dynamical MHD simulations. This unique data set allows for a more comprehensive exploration of the methods. Additionally, in this work we introduce a systematic, yet ``blind'' analysis, where we rigorously evaluate how different assumptions affect the derived quantities, without prior knowledge of the actual value of the magnetic field, in order to avoid any bias in the evaluation.

Our manuscript is organized as follows. In Sect.~{\ref{methodology}}, we describe the simulation and the data that we used. In Sect.~{\ref{analysis}}, we present the simulated cloud and describe the several approaches used to estimate the three key parameters ($\rho$, $\delta$v, $\delta\theta$). In Sect.~{\ref{estimatingB}}, we apply the two methods for all different combinations of the estimated parameters and we derive values for the strength of the POS component of the magnetic field. In Sect.~{\ref{final_comp}}, we compare our estimated magnetic field values with the ``real'' ones extracted from the simulation and quantify the accuracy of each method. In Sect.~{\ref{Discussion}}, we discuss our results and propose a methodology for observers to adopt, when dealing with self-gravitating clouds. In Sect.~{\ref{Summary}}, we summarize our results.

%--------------------------------------------------------------------
\section{Methodology}\label{methodology}

\subsection{MHD chemo-dynamical simulation}

We use a 3D simulation of a turbulent collapsing molecular cloud described in {\citet{TRITSIS2025b}}, performed using the \textsc{FLASH} Adaptive Mesh Refinement code ({\citealt{Fryxell2000, Dubey2008}}). The setup utilizes ideal MHD with non-equilibrium chemistry and an isothermal ideal gas equation of state, with a constant temperature of $\rm{T} = 10 \, \rm{K}$. The initial number density of the cloud is \( 500 \, \text{cm}^{-3} \), and the magnetic field is initialized along the $z$ axis with a strength of \( 7.5 \, \mu \text{G} \). Therefore, the initial mass-to-flux ratio in the simulation (expressed in units of the critical value for collapse, {\citeyearless{Mouschovias1976}}) was approximately 2.3. The cloud has a total mass of about \( 240 \, \rm{M_\odot} \) and spans 2 pc in each dimension, using a base grid resolution of \( 64^3 \) cells with two levels of refinement. Therefore, the effective resolution is approximately \( 8 \times 10^{-3} \, \text{pc} \). The evolution was tracked until the central number density reached \( 10^5 \, \text{cm}^{-3} \), which is also the timestamp that we analyze here. Turbulent velocity field was initialized using the public code \texttt{TurbGen} {\citep{Federrath2022}}, generating a power-law velocity spectrum \( \frac{dv^2}{dk} \propto k^{-2} \) over wavenumbers \( 2 \leq k \leq 20 \).

We model decaying turbulence in the cloud, initializing the velocity field with sonic and Alfv\'enic Mach numbers of approximately $\rm{\mathcal{M}_s = \sigma_v/c_s \approx 3}$ and $\rm{\mathcal{M}_A = \sigma_v/v_a \approx 1.2}$ respectively, where $\rm{\sigma_v}$ is the turbulent velocity dispersion, $\rm{c_s}$ is the sound speed, and $\rm{v_A}$ is the Alfv\'en velocity,  and a turbulence driving parameter of $\zeta = 0.5$. This setup corresponds to a natural mixture of modes, where 2/3 of the turbulent energy is injected into solenoidal motions and 1/3 into compressive motions (see Fig.1 of \citealt{FEDERRATH2010}). We note that these represent the initial conditions of the simulation and do not necessarily reflect the dynamical state of the cloud at the later stages, which is the focus of our analysis. Using a Helmholtz decomposition of the velocity field at those later times, we find that the fraction of solenoidal power increases to approximately 70\%. This evolution is expected, as compressive motions dissipate more rapidly than solenoidal ones (e.g., \citealt{PADOAN2016, YUE}), leading to a natural shift in the energy distribution during the decay of turbulence.

\begin{figure*}[htbp] 
    \centering
    \includegraphics[width=0.9\textwidth,height=0.7\textheight,keepaspectratio]{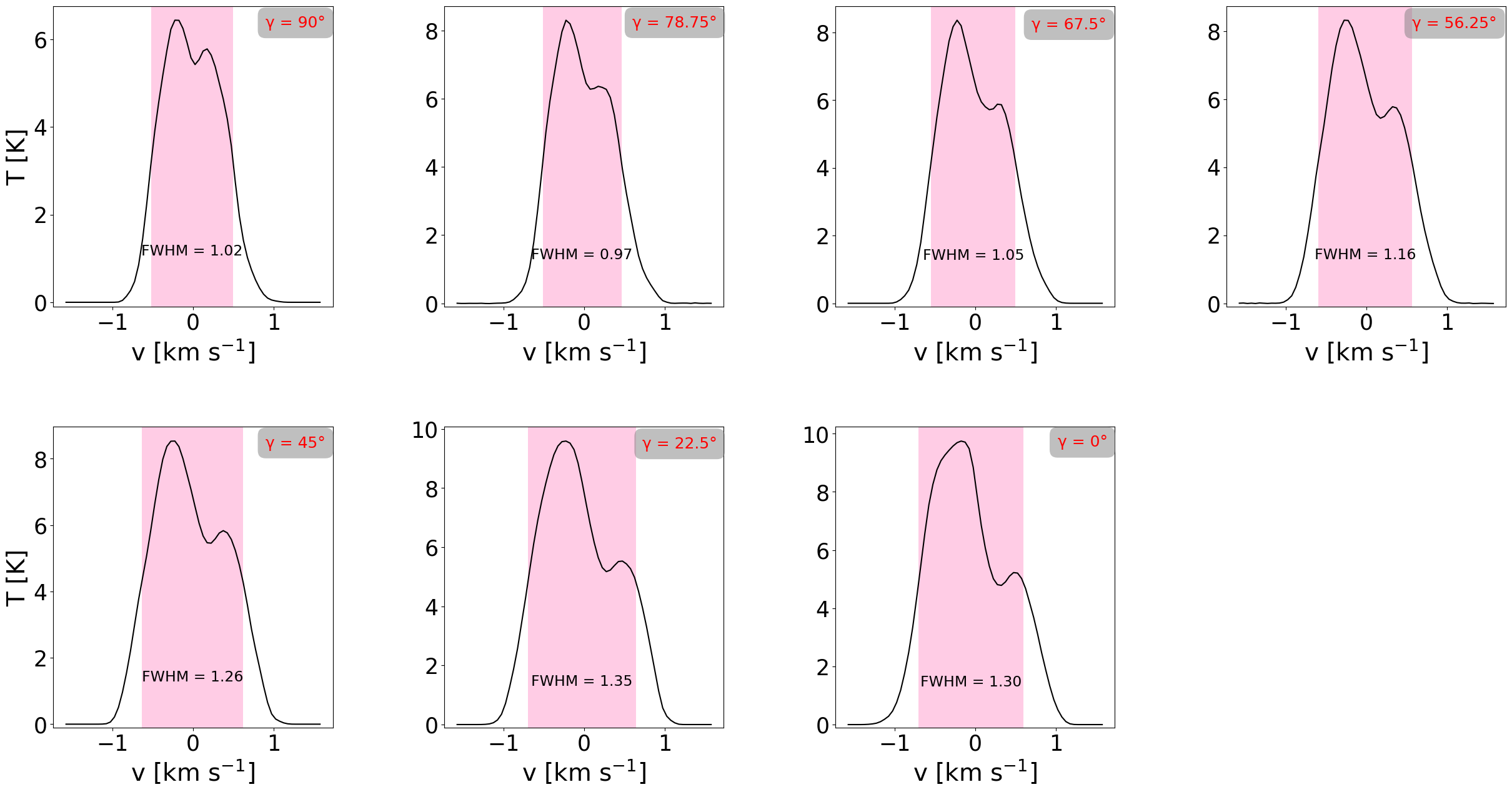}
    \caption{Average spectra and their FWHM as a function of the inclination angle $\gamma$. The rose shaded areas denote the FWHM that we calculated with the exact numerical value annotated in km $\rm{s^{-1}}$ on every panel.}
    \label{fwhm}
\end{figure*}

\subsection{Synthetic obsrevations}

We use this chemo-dynamical simulation to generate synthetic observations. Dust polarization observations were generated following the methodology of {\citet{King2019}}, with the Stokes parameters I, Q, U computed in column density units. The polarization angle is defined as

\begin{equation}
    \rm{\theta = \frac{1}{2} \arctan\left( \frac{U}{Q} \right).}
\end{equation} 

Synthetic spectral-line observations were created using the \textsc{PyRaTE} non-LTE radiative-transfer code ({\citealt{Tritsis2018, Tritsis2023a}}). Specifically we used the CO $\rm{J = 1\rightarrow0}$ and $\rm{J = 2\rightarrow1}$ lines. The ray-tracing approach in \textsc{PyRaTE} solves the radiative-transfer problem while considering that photons emitted from one region may interact with another region, provided their relative velocity is smaller than the thermal linewidth. Radiative-transfer post-processing was performed assuming five rotational levels and a grid resolution of \( 128^3 \). A spectral resolution of \( 0.05 \, \text{km/s} \) was used, with 64 points in frequency space. Noise was added to the spectra to ensure a signal-to-noise ratio (SNR) of 20 in the cloud center. Finally, we also generate column-density maps.

We used $^{12}$CO as the molecular tracer for our analysis. While other species such as NH$_3$, N$_2$H$^+$, or C$^{18}$O are commonly used to probe the denser regions of molecular clouds, $^{12}$CO is the second most abundant molecule after H$_2$ and, due to its relatively low effective critical density (e.g., \citealt{SHIRLEY2015}), it traces the more extended, lower-density regions of the cloud. As shown in Fig. 2 of \cite{TRITSIS2025a}, $^{12}$CO is significantly more widespread and abundant compared to high-density tracers. As such, it is well suited for capturing the large-scale turbulent motions that are most relevant for our analysis. Since both the DCF and ST methods are based on the assumption of equipartition between kinetic and magnetic energies---a condition that is generally valid at large spatial scales (e.g., \citealt{BEATTIE2025})---using $^{12}$CO is more appropriate. In contrast, had we used higher-density tracers such as N$_2$H$^+$, we would have biased our measurements toward compact regions. In these denser regions, the validity of the equipartition assumption between the magnetic and kinetic energies (and therefore the applicability of these methods) is dubious (see also the discussion in Appendix \ref{appendix}). Although optically thinner alternatives such as $^{13}$CO or C$^{18}$O could reduce optical depth effects while still tracing much of the cloud, they were not included in our chemical network, and we chose to avoid introducing additional assumptions about their abundances relative to CO.

Synthetic observations were generated for seven different polar inclination angles (\( \gamma = 0^\circ, 22.5^\circ, 45^\circ, 56.25^\circ, 67.5^\circ, 78.75^\circ, 90^\circ \)) with respect to the mean component of the magnetic field, which is along $z$ axis, allowing us to study the effect of inclination angle when estimating the magnetic field. To follow the nomenclature in the literature, this angle is denoted as \( \gamma \) throughout the remainder of this manuscript. In our notation, \( \gamma = 90^\circ \) represents the edge-on case, with the mean magnetic field on the POS, while \( \gamma = 0^\circ \) corresponds to the face-on case, with the mean magnetic field along the line of sight (LOS).

% , and the rest of the angles represent intermediate cases.

\section{Analysis}\label{analysis}

To ensure an unbiased approach of our study, two independent teams were established in the beginning of our analysis. The team responsible for applying the DCF and ST methods to estimate the magnetic field strength, only had access to the same quantities as an observer would have: column density and polarization maps, PPV cubes of the CO $\rm{(J = 1\rightarrow0)}$ and CO $\rm{(J = 2\rightarrow1)}$ transitions, without direct knowledge of the true magnetic field values or any other intermediate quantities. The inclination angle and the constant kinetic temperature (T=10K) were the only extra information that was provided to guide the analysis. This separation was carefully maintained throughout the analysis phase, limiting potential biases in the derivation and/or interpretation of intermediate results until we had our ``observationally-estimated'' magnetic field values that we could compare to the true values in the simulation. At the same time, while is true that the basic equations (Eq. \ref{DCF_eq}, and \ref{ST_eq}) do not depend on the inclination angle, the final observables do. Therefore, by using the inclination angle as a guide, we simultaneously study its effect on the final estimates of magnetic field strength.

We set a threshold for the column density at $10^{21} \rm{cm^{-2}}$, such that all synthetic data that fall in regions of the cloud that do not exceed this threshold, are removed from our analysis. Star formation and the effects of self-gravity are more relevant in regions with higher column densities. Setting a threshold aligns the dataset with regions where such effects are likely to be present and significant (e.g., {\citealt{TAFALLA, HACAR}}). We further explore this choice in Appendix {\ref{appendix}} where we repeat our analysis with an increased column density threshold value by one order of magnitude.

In Fig. {\ref{maps}}, we show the column density maps (left column) and the CO (J=1$\rightarrow$0) spectral line first and second moment maps of the cloud (middle and right columns, respectively) for three different inclination angles. The overplotted red segments on each panel show the polarization vectors, and are the same across each row (i.e., panels refer to the same inclination angle). As expected, the column density is maximum at the center of the cloud, where we also observe some filamentary structures. Regarding the morphology revealed by the polarization pseudovectors, the hourglass morphology is evident at $\gamma=90^\circ$. At $\gamma=45^\circ$, the magnetic field appears significantly more turbulent but a ``mean'' component can still be defined. At $\gamma=0^\circ$ the magnetic field appears completely random.

In the following sections we describe the various approaches used to estimate each of the three parameters that enter in the DCF and ST formulas.

\subsection{Derivation of the velocity dispersion $\delta$v}

The first step of our analysis focuses on estimating the velocity dispersion using the three different approaches described below. We emphasize that the data used are identical in all different approaches. To derive the $\delta$v parameter, we used only the CO (J=1$\rightarrow$0) transition. This choice is justified by the fact that the differences in the derived values between this and the CO (J=2$\rightarrow$1) line are minimal—on the order of 2\%.

\subsubsection{Derivation of $\delta$v from the average spectrum}\label{dv1}

We begin our investigation by first employing the simplest possible analysis used in the literature. The first and most straightforward approach to calculate the velocity dispersion is to derive it from the average spectrum of an observed region. Using the PPV cubes for each inclination angle, we determine the Full Width at Half Maximum (FWHM) of the average spectral line profiles. To calculate the FWHM we use spline interpolation and simply compute the width between the two points where the spectrum equals half its maximum value. In Fig. {\ref{fwhm}} we show the average spectrum as a function of the inclination angle. The rose shaded areas denote the FWHM that we calculated with the exact numerical value annotated in km $\rm{s^{-1}}$.

\begin{figure}[H] % 'H' to place the figure exactly here
    \centering
    \includegraphics[width=0.47\textwidth,height=0.3\textheight,keepaspectratio]{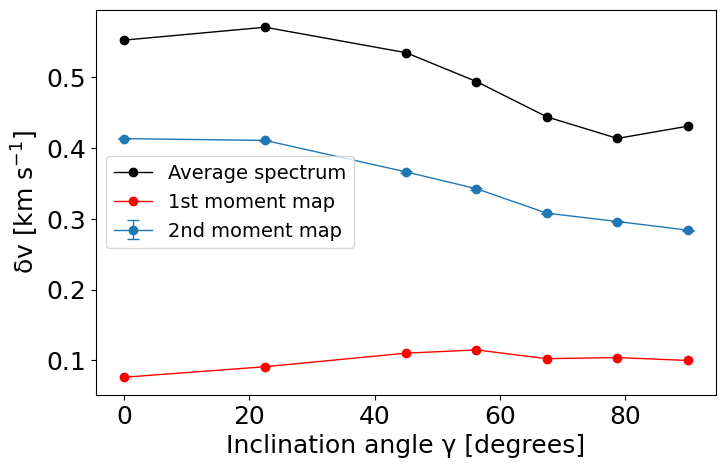} 
    \caption{$\delta$v as a function of the inclination angle ($\gamma$ = 90$^\circ$ denotes the edge-on case, with the mean magnetic field on the POS, while $\gamma$ = 0$^\circ$ denotes the face-on case, with the mean magnetic field along the LOS). The black line represents the values extracted using the average spectra, with the blue line we show the derived $\delta$v using the second moment maps, and with the red line using the first moment maps.}
    \label{dv_fmm}
\end{figure}

The results for $\delta$v as a function of the inclination angle are shown in Fig. {\ref{dv_fmm}} (black solid line). As is evident, there is a general trend that $\delta$v increases as the inclination angle approaches $\gamma=0^\circ$ (face-on), but the increase using this method is not strictly monotonic. A broader FWHM arises when observing the cloud face-on (along to the magnetic field direction) because the cloud contracts more rapidly along the mean component of the magnetic field.

\subsubsection{Derivation of $\delta$v from the second moment map}\label{dv2}

A more refined approach for determining the velocity dispersion ($\delta$v), is to use the second moment maps (e.g., {\citealt{CARMA}}). We compute the mean of the second moment map over the entire simulated cloud, for $\rm{N_{H_2} \geq 10^{21} cm^{-2}}$. We calculate the second moment maps from the PPV cubes and generate 1D distributions to constrain $\delta$v which we take it to be equal to the mean of these distributions. As the error of $\delta$v we take the standard deviation of the 1D distributions, over the number of data points.

\begin{figure*}[htbp] 
    \centering
    \includegraphics[width=0.9\textwidth,height=0.7\textheight,keepaspectratio]{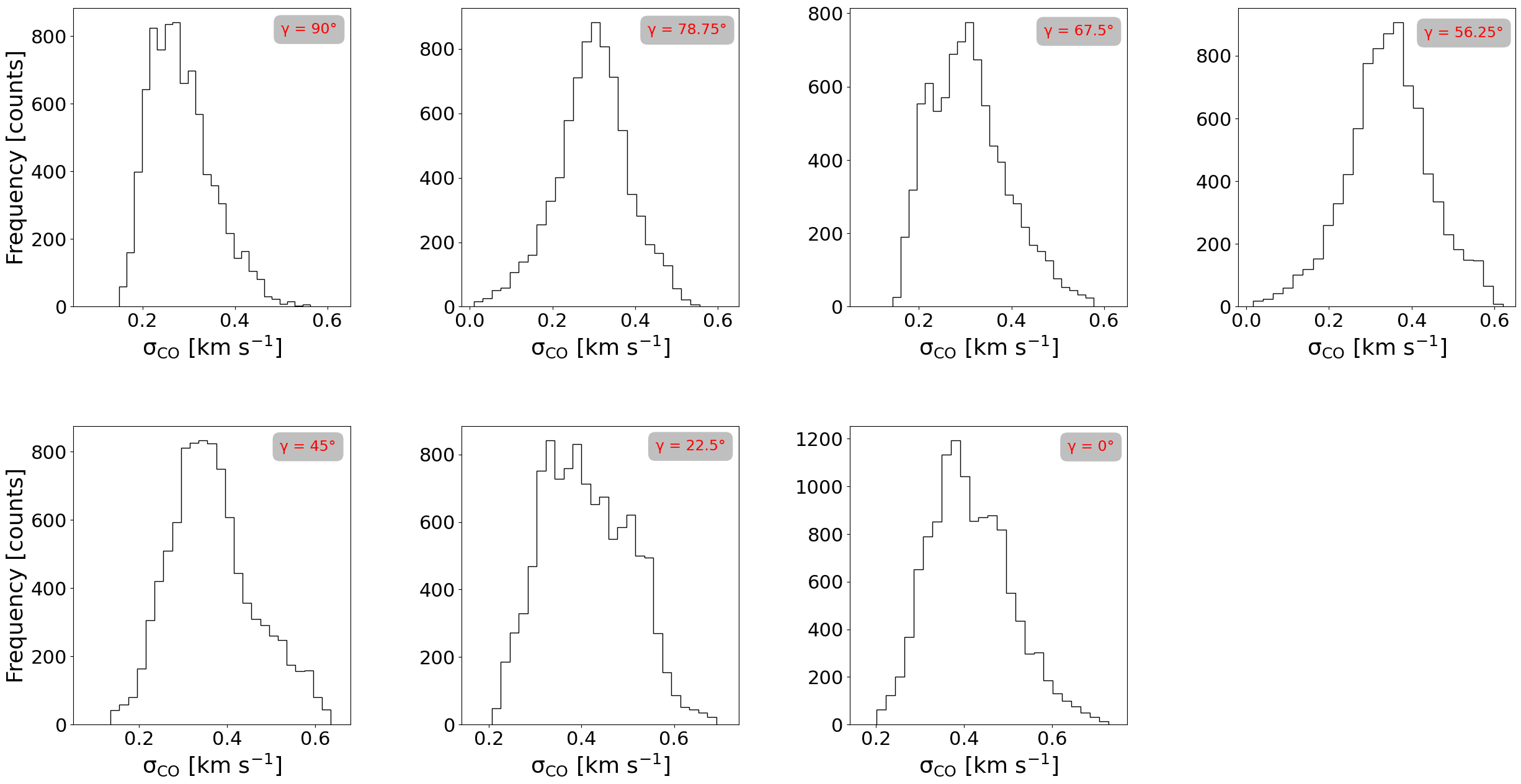}
    \caption{1D distributions of the second moment maps as a function of the inclination angle $\gamma$.}
    \label{smm_dist}
\end{figure*}

The resulting distributions for each inclination angle (\(\gamma\)) are shown in Fig. {\ref{smm_dist}}. For certain inclination angles, the resulting distributions closely resemble Gaussians, while for other angles the distributions deviate from Gaussianity. We attempted fitting both simple and skewed Gaussians, but found minimal differences in the final results, therefore we opted to use the simplest approach, which is taking the values directly from the distributions.

The results for $\delta$v as a function of the inclination angle are presented in Fig. {\ref{dv_fmm}} (blue solid line). As expected, the velocity dispersion is larger when extracted from average spectrum compared to second moment map due to the way spatial averaging smooths out local variations in velocity. When calculating the velocity dispersion from the average spectra, the line profiles are spatially averaged along the observer's LOS, covering the entire depth of the cloud. This integration combines both low - and high - velocity components within the cloud, leading to a broader velocity distribution. In contrast, second moment map captures localized velocity dispersions at each location, preserving small-scale variations and yielding smaller dispersions. 

As shown in Fig.{\ref{dv_fmm}}, the trend of the velocity dispersion derived using each of the two approaches described above with the inclination angle is very similar; however, in this case, it behaves closer to a monotonic trend. This more consistent behavior suggests that this approach is more reliable, as it better minimizes the influence of other factors discussed in Sect. {\ref{dv1}}, resulting in a more uniform and predictable trend.

\subsubsection{Derivation of $\delta$v from the first moment map}\label{dv3}

In the limit where turbulence is isotropic the velocity dispersion should be constant with inclination. Variations of the velocity dispersion with the inclination angle are probably due to gravity. To address this
limitation, our next step involves utilizing the first moment maps to estimate the velocity dispersion (e.g., {\citealt{CARMA, Stewart2021}}).

To analyze the central regions of the maps, we first note that for intermediate angles, the vertical POS dimension extends beyond the range of -1 to 1 pc. We decide to restrict to the interval $[-1, 1]$. By adopting this procedure, we isolate the core regions of interest. Subsequently, we generate gradients for each inclination angle by smoothing the first moment maps, using a kernel size equal to half the box size. The smoothed maps are then subtracted from the original maps, effectively removing large-scale gradients that are unlikely to be associated with the turbulent motions we aim to constrain. We then compute the velocity dispersion, $\delta$v, by calculating the standard deviation of the large-scale gradient-subtracted first moment maps. The red curve in Fig. {\ref{dv_fmm}} show the velocity values obtained with this approach.

The results for the three approaches that we used to estimate $\delta$v are summarized in Fig. {\ref{dv_fmm}}. As expected, the values from the first moment maps are significantly lower than those extracted from the second moment maps or the average spectra. Depending on the methodology used to derive $\delta$v we can get differences up to factors of 5, whereas different approaches have somewhat different dependence with the inclination angle.

\subsection{Estimation of the polarization angle distribution dispersion $\delta\theta$}

The second step of our analysis focuses on quantifying the spread in the polarization angle distributions. Similar to the velocity dispersion ($\delta$v), we explore multiple approaches for calculating this quantity, though in this case we focus on two distinct approaches. The two approaches are detailed in the following sections.

\subsubsection{Deriving $\delta\theta$ from the spread of the polarization angle distribution}\label{Pol0}

\begin{figure*}[htbp]
    \centering
    \includegraphics[width=0.9\textwidth,height=0.7\textheight,keepaspectratio]{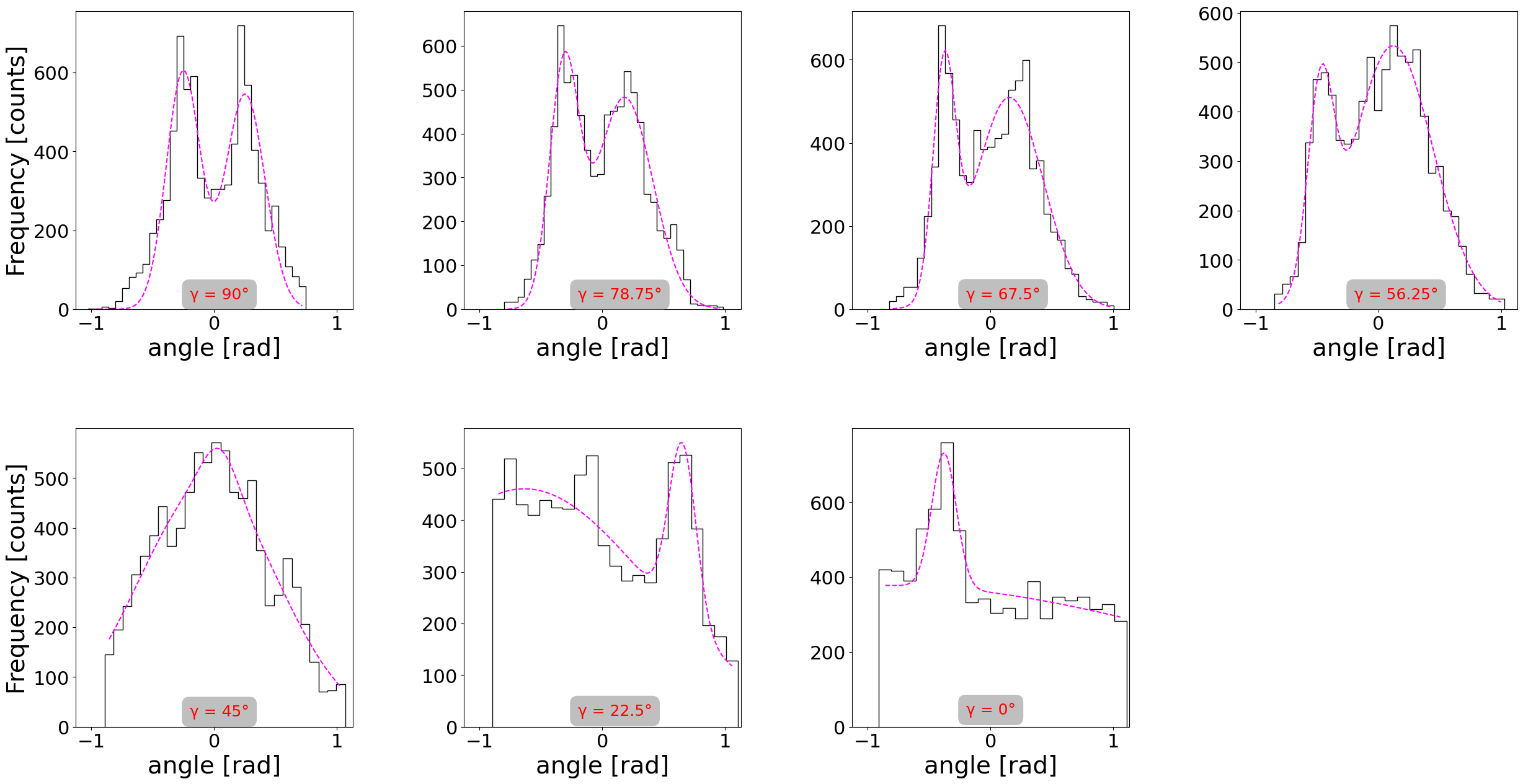}
    \caption{The black solid lines show the polarization angle distributions for every inclination angle $\gamma$, while the magenta dashed lines indicate the Gaussian fits applied to these distributions. For $\gamma \geq 56.25^\circ$, the distributions exhibit a distinct bimodality that disappears as $\gamma$ decreases.}
    \label{dtheta_fits}
\end{figure*}

As a zeroth-order approximation, we calculate $\delta\theta$ by simply taking the circular standard deviation of the distributions of the polarization angles. These distributions are shown in Fig. {\ref{dtheta_fits}}. The magenta dashed lines represent fits to these distributions and are explained in Sect.{\ref{dtheta2}}. For the analysis considered in this section, the results from the fitting are not used in any shape or form. For $\gamma \geq 56.25^\circ$, the polarization angle distributions are bimodal. This bimodality is consistent with the hourglass morphology of the magnetic field, as evidenced by the pseudo-vectors shown in Fig.~{\ref{maps}}. As the cloud is observed closer to the face-on case ($\gamma=0^\circ$), the hourglass morphology is no longer observable and the bimodality in the distributions vanishes. The results for $\delta\theta$ with this approach as a function of inclination angle are shown in Fig. {\ref{dtheta_comp_fits}} (purple dashed line).

\subsubsection{Deriving $\delta\theta$ through Gaussian fitting on the polarization angle distribution}\label{dtheta2}

Given the initial conditions of our simulation, the hourglass morphology naturally emerges because of the collapse perpendicular to the lines after the initial thermal relaxation. This can then be observed for certain viewing angles. As shown in Fig. \ref{maps}, a clear hourglass shape is visible when the cloud is viewed edge-on (i.e., at an inclination angle $\gamma = 90^\circ$). Hourglass morphologies have been observed in protostellar cores (e.g., \citealt{GIRART2006, FRAU2011}), but not in starless or prestellar ones. The presence and observability of such features depend on several factors, including the orientation of the magnetic field with respect to the line of sight, as demonstrated in our results (see Fig. \ref{dtheta_fits}), as well as observational limitations. If such a morphology is observed, its contribution to the dispersion of polarization angles can be removed using the method described in this section. This work does not aim to assess whether these morphologies exist in starless or prestellar cores, but rather focuses on how to eliminate their influence on the polarization angle dispersion when they are present and detectable.

To improve upon the previous method (Sect. \ref{Pol0}), we need to remove the hourglass morphology that we observe in the polarization lines, because both methods (DCF and ST) omit the effect of self-gravity. We expect this bimodality in the distributions to statistically increase the spread, as the separation between the peaks introduces additional variability compared to a unimodal distribution. We perform an analytical fit considering a linear combination of two Gaussian functions (e.g., \citealt{PALAU2021}), expressed as

\begin{equation}
    f(x) = \sum_{i=1}^{2} A_i \cdot \frac{1}{\sigma_i \sqrt{2 \pi}} \exp\left( - \frac{(x - \mu_i)^2}{2 \sigma_i^2} \right),
\label{eq:6}
\end{equation}
where $A_i$ is the amplitude of each Gaussian component, $\sigma_i$ is the respective standard deviation, and $\mu_i$ is the mean.

The final results of the fitting process could be significantly influenced by the number of bins and/or the chosen bin width. To mitigate such biases, we adopt a quantitative method for selecting the bin width, aiming to eliminate these biases and ensure that the fitting results are robust and independent of arbitrary binning parameters. Specifically, we use the method developed by {\cite{SCOTT}} to obtain the optimal number of bins for these distributions, since it adjusts the bin width based on both the standard deviation and the size of the dataset. The bin width is calculated using the formula 
\begin{equation}
\Delta = \frac{3.5 \sigma}{n^{1/3}},
\end{equation}
where \( \sigma \) is the standard deviation of the data, and \( n \) is the number of data points. Polarization angle distributions typically exhibit some degree of variability, and Scott's rule helps to capture this by dynamically adjusting the bin size to reflect the underlying structure of the data.

In Fig. {\ref{dtheta_fits}} we present the results of these fits for each inclination angle $\gamma$. From the fitted parameters, we determined the total spread ($\delta\theta$), as
\begin{equation}\label{quadrature}
\rm{\delta\theta = \{A_1 \cdot \sigma_1^2 + A_2 \cdot \sigma_2^2\}^{1/2}},
\end{equation}
where $A_i$ represents the normalized amplitude of each Gaussian ($\rm{A_1 + A_2 = 1}$). The associated uncertainties are calculated using a Gaussian error propagation based on Eq.~\ref{quadrature}. We add the two component widths in quadrature since it is the variances—rather than the standard deviations—that combine when independent fluctuations are mixed. Each fitted Gaussian represents an independent mode of dispersion, and the total variance of the combined distribution is simply the weighted sum of the individual variances. Taking the square root of that sum then gives the correct overall standard deviation. In contrast, a straight weighted average of the sigmas does not preserve the second moment (variance) of the mixture and systematically underestimates the true spread.

\begin{figure}[H] % 'H' to place the figure exactly here
    \centering
    \includegraphics[width=0.5\textwidth,height=0.3\textheight,keepaspectratio]{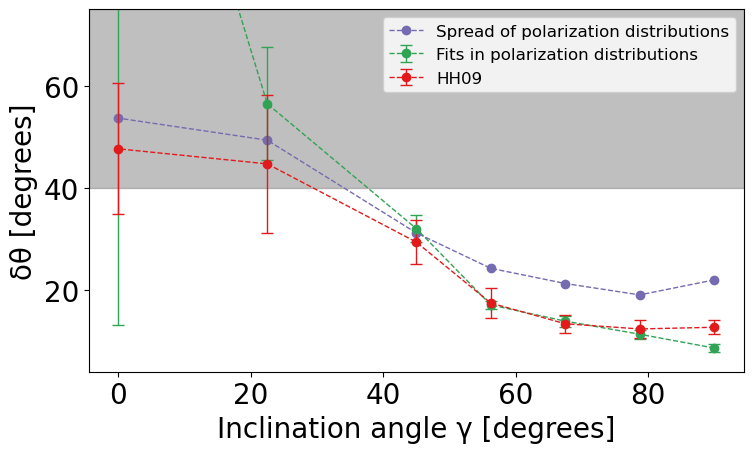} 
    \caption{$\delta\theta$ as a function of the inclination angle $\gamma$. The purple dashed line represents the values extracted using the circular standard deviation of the polarization distributions, the green dashed line using the Gaussian fits, and the red dashed line using the HH09 method. The green point at $\gamma=0^\circ$ is off the scale at value $\delta\theta\approx(150\pm140)^\circ$. The gray shaded region corresponds to values of $\delta\theta$ exceeding $\rm{40^\circ}$. The employment of the dispersion angle methods becomes questionable under such considerable spreads, because of the $\pi$ ambiguity in $\delta\theta$ which limits the dynamic range of the observable.}
    \label{dtheta_comp_fits}
\end{figure}

In Fig. {\ref{dtheta_comp_fits}}, we present the results for \(\delta\theta\) as a function of the inclination angle \(\gamma\), obtained through fitting (green dashed line), and compare these findings to \(\delta\theta\) derived directly from the circular standard deviation of the distributions (purple dashed line).

The observed results reveal notable trends. For inclination angles greater than \(45^\circ\), the values of \(\delta\theta\) are lower than those obtained previously, while the overall trend remains consistent, with the sole but very important exception being the final angle ($\gamma = 90^\circ$). For $\gamma=45^\circ$, the values of $\delta\theta$ derived from the different approaches appear to be very close. This is expected, as the polarization angle distribution for this angle resembles a single Gaussian (lower left panel of Fig. {\ref{dtheta_fits}}), meaning that removing the hourglass morphology is not particularly significant in this case. For inclination angles below \(45^\circ\), the uncertainties in the measurements increase significantly, with the angle \(\gamma = 0^\circ\) exhibiting particularly substantial errors. This stands to reason, since the distributions for these inclination angles are not close to a Gaussian, or a combination of Gaussians.

Similarly to $\delta$v, we also tested skewed Gaussian fits for $\delta\theta$ and they showed little to no difference in the final results. This suggests that the method is robust to deviations from perfect Gaussian shapes, and the choice of a simple Gaussian model is sufficient for the analysis. The shaded region in Fig.~\ref{dtheta_comp_fits} corresponds to values of \(\delta\theta\) exceeding \(40^\circ\), indicating that the reliability of the DCF and ST methods becomes questionable under such considerable spreads, since the mean component of the magnetic field is not well defined.

\subsubsection{Deriving $\delta\theta$ through a dispersion function}\label{dtheta3}

In this subsection, we implement the method proposed by \cite{HOUDE}, hereafter HH09\footnote{The implementation of the HH09 method was performed after the revision process, meaning that the team responsible for processing the mock observations already had knowledge of the true magnetic field strength values. However, given the fact that we do not proceed to derive any $\rm{B_{POS}}$ values using this implementation there are no unconscious biases involved in our final estimates of $\rm{B_{POS}}$.}. This method involves the calculation of the dispersion function of the polarization angles, and subsequently fitting an analytical expression in order to extract the ratio of the turbulent to ordered components of the magnetic field. This method is physically motivated as the expressions used to fit the dispersion function are analytically derived. At the same time however, it involves several methodological choices (see Appendix \ref{appendix2} for more details) which can affect the final results.

The derived values of $\delta\theta$ using the dispersion function are shown in Fig. \ref{dtheta_comp_fits} as red points. As illustrated, the values obtained from this method are in strong agreement with those derived through Gaussian fitting across all inclination angles. The only exception is for $\gamma$ = 90$^\circ$ (edge-on case), where the two sets of values do not align within the error bars. In the interest of simplicity and to avoid introducing additional free parameters, we adopt the Gaussian-fitting approach for the remainder of our analysis. However, we note that while the Gaussian-fitting method effectively removes the hourglass morphology, the HH09 method could offer a more robust estimate of $\delta\theta$ in cases where the large-scale magnetic field morphology differs from an hourglass. This alternative approach might better account for variations in the polarization angle distribution that the Gaussian method does not fully capture.

\subsection{Estimation of the cloud density $\rho$}\label{Dens}

To estimate the cloud density we use three distinct approaches. The first two are geometric approaches that include two different assumptions regarding the cloud's shape and are outlined in the first two following sections. The third approach, described in Sect~{\ref{Dens3}}, involves inverse radiative transfer analysis, which introduces additional complexity and requires supplementary data. It is crucial to note that no information about the 3D shape of the cloud has been disclosed to the team applying the magnetic field strength estimation methods.

\subsubsection{First geometrical approximation: filamentary approach}\label{Dens1}

To estimate the number density from the column density, we firstly assume the cloud has a cylindrical shape (e.g., {\citealt{ANDRE, ARZOUMANIAN}}), with LOS dimension equal to \(0.1 \, \text{pc}\). Thus, the number density, $\rm{n_{\mathrm{H}_2}}$, is given by \(\rm{n_{\mathrm{H}_2} = \rm{\overline{N}_{H_2}}/0.1[pc]}\), where $\rm{\overline{{N}}_{H_2}}$ represents the mean column density.

\subsubsection{Second geometrical approximation: spherical approach}\label{Dens2}

\begin{figure}[h] % 't' for top placement
    \centering
    \includegraphics[width=0.5\textwidth,height=0.3\textheight,keepaspectratio]{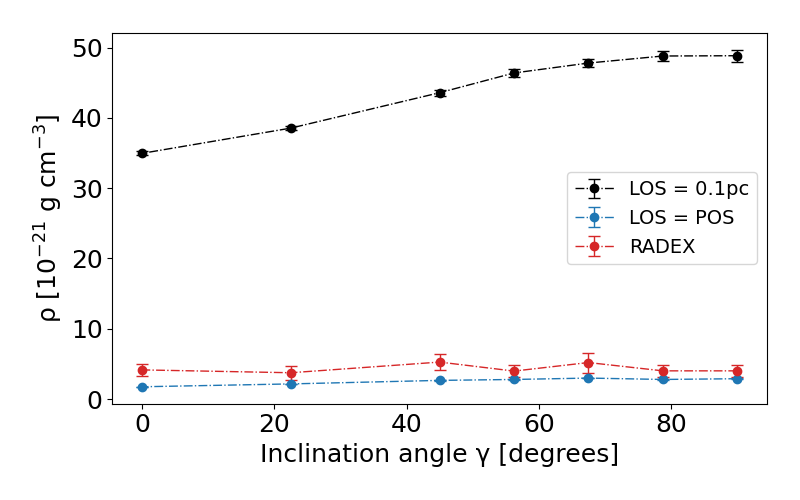} 
    \caption{Density as a function of the inclination angle $\gamma$. The black line represents the values extracted using the filamentary approach, the blue line with the spherical approach, and the red line using RADEX.}
    \label{comp_densities}
\end{figure}

In order to make a different, but still geometrical, estimation of the density, we decide to set the dimension of the cloud along the LOS to be comparable to its POS dimension. We first calculate the average radius in pixel units by applying our column density threshold (see Fig. \ref{maps}) to define the cloud boundary, measuring the distance between the minimum and maximum pixel positions along two orthogonal axes, and averaging those spans. This mean pixel radius is then converted into a physical length using the underlying physical resolution. Using this estimate for the dimensions of the cloud we then calculate the number density as \(\rm{n_{\mathrm{H}_2}} = \rm{\overline{N}_{H_2}}/\rm{R_{POS}}[pc]\), where $\rm{R_{POS}}$ is the average radius that we visually calculated. In Fig. {\ref{comp_densities}}, we show the results of the densities calculated by the two different methods (black and blue dashed-dotted lines, respectively).

\subsubsection{Radiative transfer analysis}\label{Dens3}

A more detailed approach used in the literature for deriving the density is by performing an inverse radiative-trasnfer analysis with the radiative-transfer code \textsc{RADEX} ({\citealt{RADEX}}). \textsc{RADEX} takes as input the collision partner ($\rm{H_2}$ in our case), the frequency range, the kinetic temperature of the cloud{\footnote{It is important to note that we already knew the constant kinetic temperature throughout the cloud ($\rm{T=10K}$). If the temperature were unknown, it could introduce additional degeneracies in the density calculation, which are not addressed in this study.}}, the number density of the collision partner, the column density of the molecule, and the FWHM of the line. It outputs key properties for each transition, such as the line optical depth, defined as the optical depth of the equivalent rectangular line shape and the line intensity (in antenna temperature units) at the rest frequency of each line.

The methodology that we use to find the \( \rm{n_{\text{H}_2}} \) from RADEX is the following. The data that we use, are the PPV cubes for the 1\(\rightarrow\)0 and the 2\(\rightarrow\)1 transitions of CO. For the line width parameter (FWHM) we extract the second moment map for every transition line separately (and every inclination angle) and compute the mean. For the final FWHM value, we calculate the mean of the two transition lines. We then create a grid of ten values for the \( \rm{n_{H_2}} \) where the upper/lower limits are the values that we found from the previous simple methods for estimating the density (LOS = POS and LOS = 0.1 pc), and ten values for the CO column density ($\rm{N_{CO}}$) calculated from reasonable observational limits of the abundance of CO in the cloud ($\rm{X_{CO}}$ = $5\times10^{-7}$ to $7\times10^{-5}$, e.g., \citealt{SHEFFER, TRITS}). The chosen values for $\rm{n_{H_2}}$ lie within the range of $10^2$ to $10^4$ $\rm{cm^{-3}}$. Therefore, we create a grid of values with 100 different combinations used as input to \textsc{RADEX} in order to calculate the expected line intensity. We then multiply the absolute values of each temperature by 1.06 and the mean of the second moment map of the line to obtain a value approximating the mean of the zeroth moment map. We additionally calculate the ratio between the zeroth moment maps of the two transitions. To determine the $\rm{n_{H_2}}$, $\rm{N_{CO}}$ combination that best reproduces the intensities of the synthetic CO emission lines, we follow the following steps. For each combination, we calculate the antenna temperature ratio of the two CO lines and set a maximum error of approximately 0.5. Among these, we select the combination for which the individual temperatures of each transition calculated with \textsc{RADEX} are closest to the synthetic ``observed'' values. As an error to $\rm{n_{H_2}}$ we take half of the ``distance'' from the next (or previous) value of \( \rm{n_{H_2}} \) in the grid that gives the closest possible ratio to the expected one.

Inverse radiative transfer analysis has been commonly employed in the literature to estimate the density of ISM clouds, particularly for estimating the magnetic field strength using the DCF or ST methods (e.g., \citealt{SANTOS2016, BONNE2020, BESLIC2024}). In Fig. {\ref{comp_densities}}, we show the results of the densities calculated with this new approach (red dashed-dotted line) compared to the previous assumptions.

\section{Estimation of the magnetic field strength}\label{estimatingB}

Now that we have calculated each parameter using multiple approaches, we proceed to apply the DCF and ST methods using every possible combination between the different approaches. In total we have 18 different combinations. This section is divided into two sub-sections, each dedicated to one of the estimation methods.

\begin{figure*}[p]  % Use 'p' to place figure on a dedicated page if necessary
    \centering
    \begin{subfigure}{\textwidth}
        \centering
        \includegraphics[width=\textwidth,height=0.42\textheight,keepaspectratio]{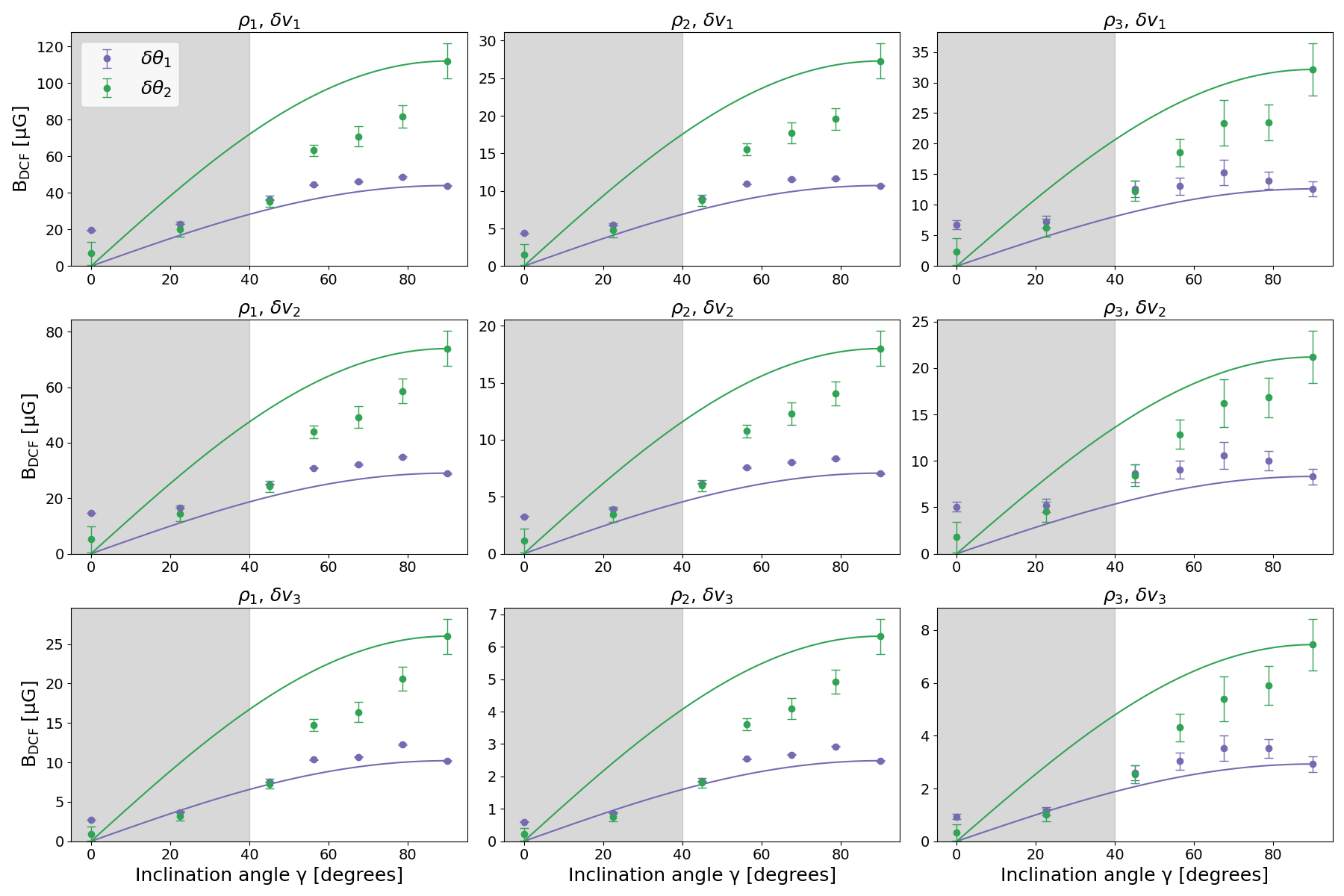}
        \caption{DCF (f=0.5) method}
        \label{DCF_comp}
    \end{subfigure}
    
    \vspace{0.3cm} % Adjust vertical space as needed

    \begin{subfigure}{\textwidth}
        \centering
        \includegraphics[width=\textwidth,height=0.41\textheight,keepaspectratio]{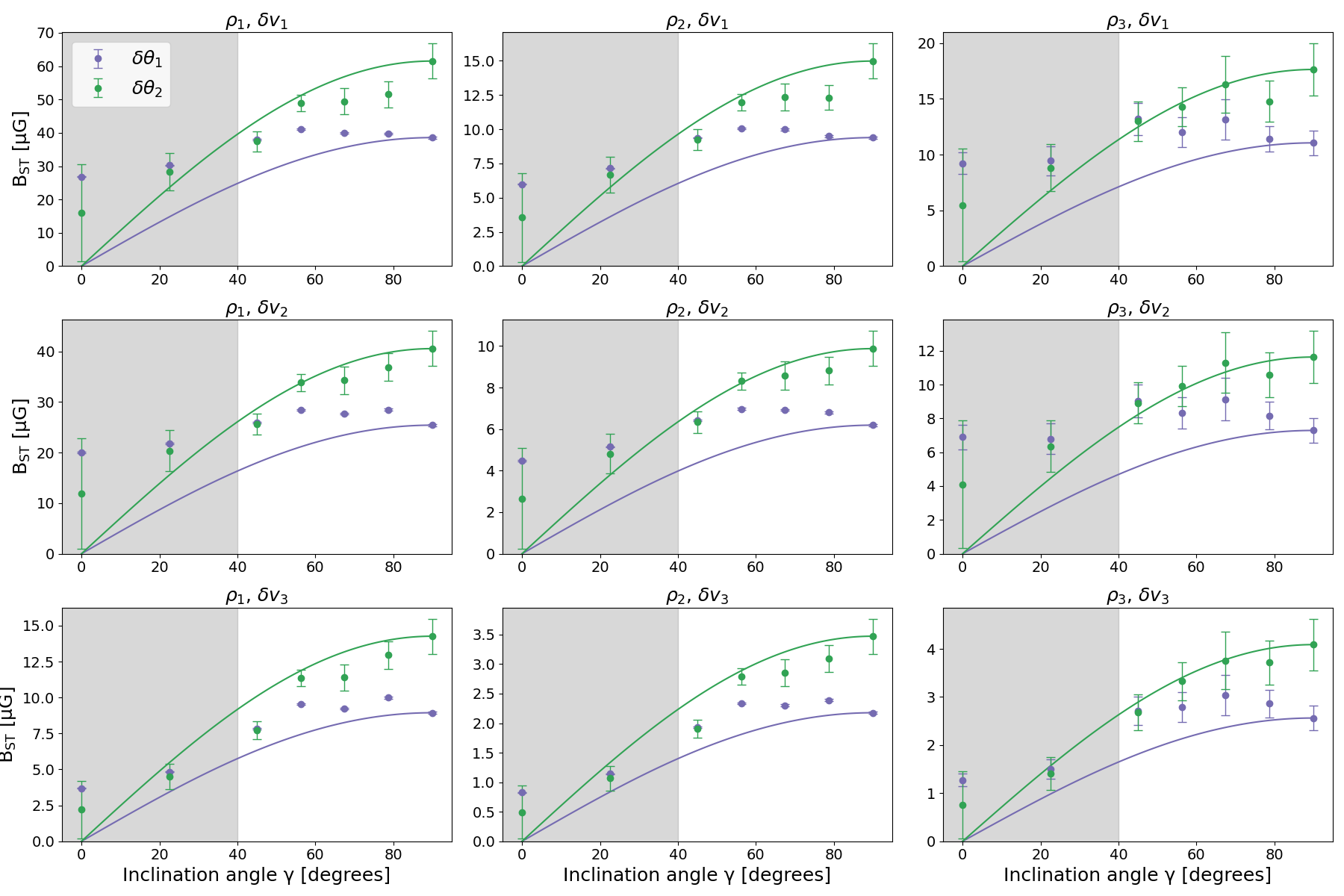}
        \caption{ST method}
        \label{ST_comp}
    \end{subfigure}

    \caption{The magnetic field strength in the POS calculated using the DCF (panels a) and ST (panels b) methods, as a function of inclination angle $\gamma$. Each individual panel represents a different combination of the parameters $\rho$, $\delta$v, $\delta\theta$, that produce different values for $\rm{B_{POS}}$. In every panel we use a different way of calculating the first two parameters, and both the ways that we estimate the last one, are denoted with the different colors. The solid lines denote the cosine decrease we expect, normalized to the estimated value of $\rm{B_{POS}}$ for an inclination angle of $\gamma = 90^\circ$. The notation of the indices that we use is explained in Table ~{\ref{table:matrix_representation}}.}
    \label{fig:combined_comp}
\end{figure*}

\begin{figure*}[h!]
    \centering
    \includegraphics[width=1\textwidth,height=0.6\textheight,keepaspectratio]{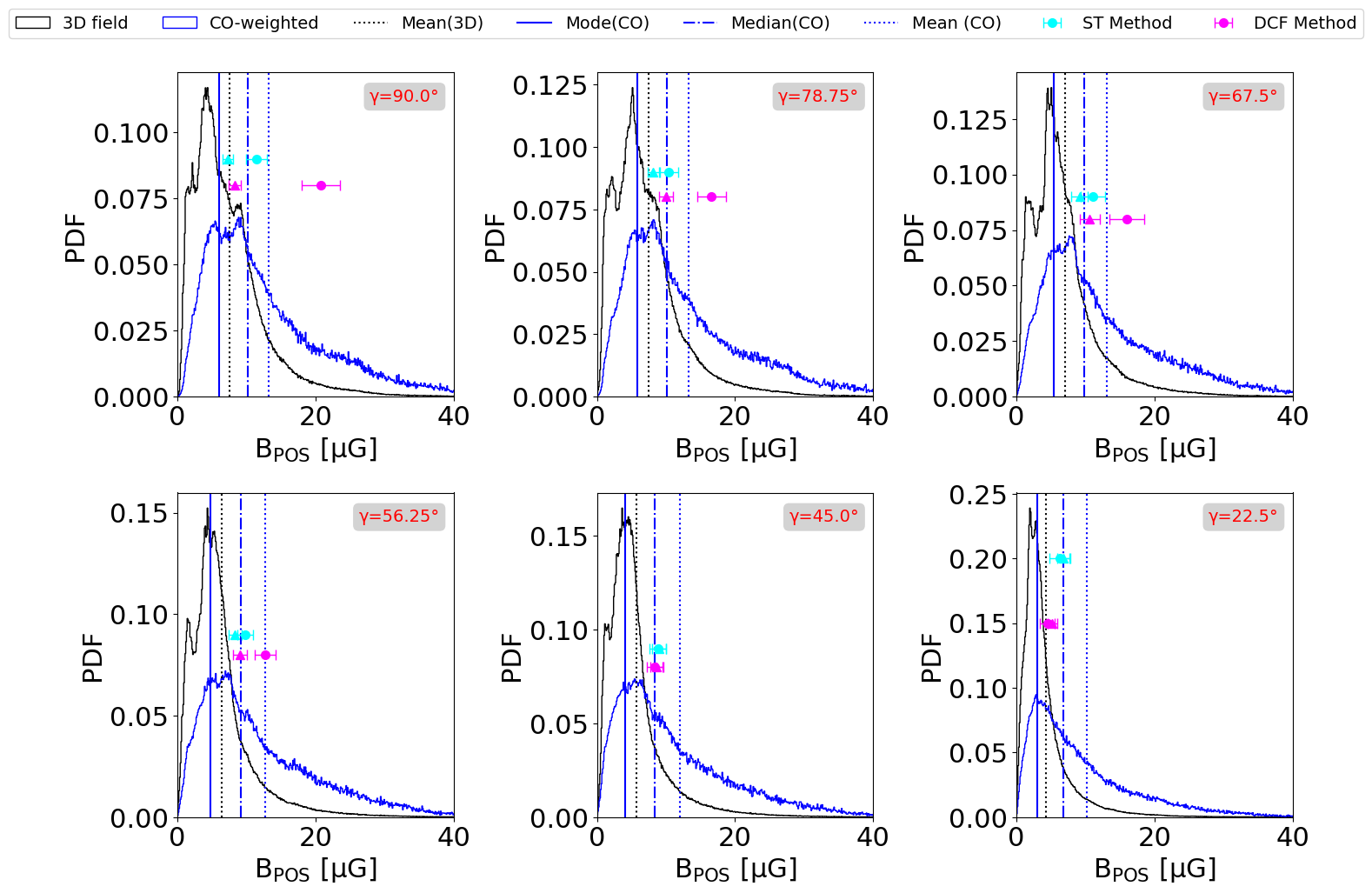}
    \caption{Distributions of $\rm{B_{POS}}$ from the simulation for different inclination angles $\gamma$. For each inclination angle we have only considered the locations where the column density exceeds the determined threshold ($10^{21} \rm{cm^{-2}}$). Black distributions show the 3D values of the $\rm{B_{POS}}$. Blue distributions show the CO-weighted values of $\rm{B_{POS}}$. Solid, dash-dotted and dotted lines mark the mode, median and mean of each distribution respectively. Points show the values that the methods we are testing predicted. The circles represent the $\rm{B_{POS}}$ values predicted by the best-performing combination, while the triangles correspond to the predictions from the second-best combination.
}
    \label{comp_lognormal}
\end{figure*}

\subsection{DCF method}

The results for the DCF (assuming f = 0.5) method are shown in Fig.~{\ref{DCF_comp}} for every inclination angle studied here. With the dots in each panel we show the derived values for the magnetic field strength. The errorbars for each data point were calculated using Gaussian error propagation, based on the uncertainties of the corresponding parameters. With the purple points we show the values of $\rm{B_{POS}}$ derived using our results for $\delta\theta$ by just considering the circular standard deviation (Section [{\ref{Pol0}}]) of the distribution shown in Fig.~{\ref{dtheta_fits}}. With the green points, we show the values of $\rm{B_{POS}}$ derived using our results for $\delta\theta$ by the Gaussian fits in the polarization angles distribution (Section [{\ref{dtheta2}}]). In the left/middle and right columns we have used, respectively, our estimates for $\rho$ assuming that the cloud is cylindrical/spherical and from our radiative-transfer analysis (Sections [{\ref{Dens1}}], [{\ref{Dens2}}], [{\ref{Dens3}}]). Finally in the upper, middle and lower rows we have used, respectively, our estimates for $\delta$v by using the average spectrum, second moment map and first moment map (Sections [{\ref{dv1}}], [{\ref{dv2}}], [{\ref{dv3}}]). The solid lines denote the cosine decrease we expect, normalized to the estimated value of $\rm{B_{POS}}$ for an inclination angle of $\gamma = 90^\circ$. The grey area stands for $\gamma<40^\circ$ where both the methods will likely not produce robust estimates.

Since the actual values of $\rm{B_{POS}}$ are not yet disclosed (see Sect.~{\ref{final_comp}}), the agreement between the points and the cosine trend is the only criterion available for comparing the different parameter combinations and assessing their performance. Although it is difficult to definitively identify which combination aligns best with the expected trend, none appear to follow it precisely. In order to facilitate the readability of the results shown in Fig.~{\ref{fig:combined_comp}}, we present in Table ~{\ref{table:matrix_representation}}, the notation that we follow.

\begin{table}[H]
\centering
\begin{tabular}{c c c c}
\hline\hline
     & \text{1} & \text{2} & \text{3} \\ \hline
\text{$\rho$}       & LOS=0.1pc & LOS=POS & RADEX  \\ 
\text{$\delta$v}    & average spectrum & $\rm{2^{nd}}$ moment map & $\rm{1^{st}}$ moment map  \\ 
\text{$\delta\theta$} & circular st. deviation & Gaussian fits & -  \\ \hline
\end{tabular}
\caption{Representation with parameters $\rho$, $\delta v$, and $\delta \theta$, and indices 1, 2, and 3, which they are used for the notation in the next figures (e.g., when $\rho$ is calculated according to the assumption LOS=POS we are using the notation $\rm{\rho_2}$, and so on). }
\label{table:matrix_representation}
\end{table}

The combination where the density is computed from inverse radiative-transfer analysis, $\delta\text{v}$ from the second moment maps and $\delta\theta$ from fitting Gaussians to remove the hourglass morphology ($\delta\text{v}_2$, $\rho_3$, $\delta\theta_2$; green points in the second row and third column of Fig.~{\ref{DCF_comp}}) appears to follow the expected trend more closely, though it also has the larger associated errors. The trends are almost the same for most of the combinations but the actual values are shifting. This stands for the lines with the same $\delta\theta$ estimation. We observe that this factor has the most significant impact on the trend of the magnetic field values. This is evident in Fig.~{\ref{fig:combined_comp}}, where the points with different colors clearly follow distinct trends, while points of the same color exhibit a more consistent pattern.

\subsection{ST method}

The results for the ST method are shown in Fig.~{\ref{ST_comp}}. Using this method for calculating magnetic field strength, we obtain values of the magnetic field strength for each of the seven inclination angles. The expected cosine trend of $\rm{B_{POS}}$ as a function of the inclination angle \( \gamma \) is the reference for comparing estimates across different combinations of $\delta$v, $\delta\theta$ and $\rho$. We observe that the green points lie close to the cosine curves, indicating that the parameter \( \delta \theta \) has the strongest influence on the trend with the inclination angle. Thus, using fits to remove the hourglass morphology (see Sect.~{\ref{dtheta2}}) better produces the anticipated trend with inclination angle. Two panels, specifically ($\delta\text{v}_2$, $\rho_3$, $\delta\theta_2$) and ($\delta\text{v}_3$, $\rho_3$, $\delta\theta_2$), appear to follow this trend well for angles \( \gamma > 40^\circ \), though the actual $\rm{B_{POS}}$ values differ by approximately half of an order of magnitude. Therefore, to select between these options, we must compare them with the true magnetic field values for each inclination angle, as we discuss in the following section.

\section{Comparison with the ``true'' magnetic field values}\label{final_comp}

At this stage of our study, the actual magnetic field values from the simulation were disclosed to the team members responsible for the analysis of the synthetic observables. The values of $\rm{B_{POS}}$ directly from the simulation were provided for each inclination angle and are shown in Fig.~{\ref{comp_lognormal}}, where different panels denote different inclination angles, enabling a direct comparison with the values we derived based on synthetic observations. As in our synthetic observations, all regions with a lower column density than the threshold discussed in Sect.~\ref{analysis} were excluded from our analysis of the ``true'' magnetic field values as well. The black distributions denote the 3D ``true'' values of the POS magnetic field and the blue lines denote the CO-weighted values of $\rm{B_{POS}}${\footnote{The distribution of the $\rm{H_2}$-weighted $\rm{B_{POS}}$ values is quite similar to the distribution of the CO-weighted $\rm{B_{POS}}$ values and is therefore not plotted in Fig.~{\ref{comp_lognormal}} to avoid cluttering.}}. The points with error bars represent the values of $\rm{B_{POS}}$ estimated using each method, keeping the estimates that are closest to the true values. The circles represent the magnetic field values predicted by the best-performing combination, while the triangles correspond to the predictions from the second-best combination.

As shown in Fig.~{\ref{comp_lognormal}}, the two methods exhibit intriguing behavior when evaluated using the same parameter combinations. Considering the cases for \(\gamma \geq 45^\circ\), where a mean magnetic field component is present, we observe that starting at \(\gamma = 90^\circ\), the ST method provides a value that closely aligns with the median of the CO-weighted distribution. DCF overestimates the $\rm{B_{POS}}$ value in this case. As we move to lower inclination angles, ST consistently provides accurate estimates, while DCF continues to overestimate. However, DCF increasingly converges toward the median, and at \(\gamma = 45^\circ\), both methods yield nearly identical values. It is noteworthy that even at \(\gamma = 22.5^\circ\), the ST method provides a remarkably accurate prediction, despite the lack of a well-defined mean magnetic field component, which makes such precision unexpected.

Although the triangle point, representing the DCF, in Fig.~{\ref{comp_lognormal}}, which indicates the best combination for the DCF, appears quite close to the median of the blue distribution, it falls noticeably outside the 1$\sigma$ range for $\gamma = 90^\circ$. The same holds true for $\gamma = 22.5^\circ$, where the ST method remains consistent.

In order to find which one of the 18 different combinations of parameters better estimates the actual magnetic field values, we calculate the $\chi^2$, only considering inclination angles $\gamma \geq 45^o$, where we expect the methods to make robust estimations\footnote{As shown in Eq. \ref{eq12} the $\chi^2$ metric is calculated for each combination of the parameters $\rho$, $\delta\theta$, and $\delta$v, after summarizing over all the inclination angles greater than 45$^\circ$. As such the performance of each combination is evaluated across all the inclination angles as a whole.}. The formula we use is
\begin{equation}\label{eq12}
\rm{\chi^2} = \sum_{\gamma} \frac{\{B_{true,\gamma} - B_{est,\gamma}\}^2}{\sigma_{est,\gamma}^2},
\end{equation}
where $\rm{B_{est,\gamma}}$ represents the estimated magnetic field strength for a specific inclination angle $\gamma$, as determined by the methods, $\rm{B_{true,\gamma}}$ is the actual magnetic field value in the simulation, and $\rm{\sigma_{est,\gamma}}$ is the uncertainty of the $\rm{B_{est,\gamma}}$ value. As the true $\rm{B_{true,\gamma}}$ value we use the median of the CO-weighted distributions that we show in Fig.~{\ref{comp_lognormal}}.

\begin{figure}[H]
    \centering  % Center the entire figure
    \begin{subfigure}{0.45\textwidth}  % Set the subfigure width (adjust as needed)
        \centering  % Center the subfigure
        \includegraphics[width=\textwidth, height=0.4\textheight, keepaspectratio]{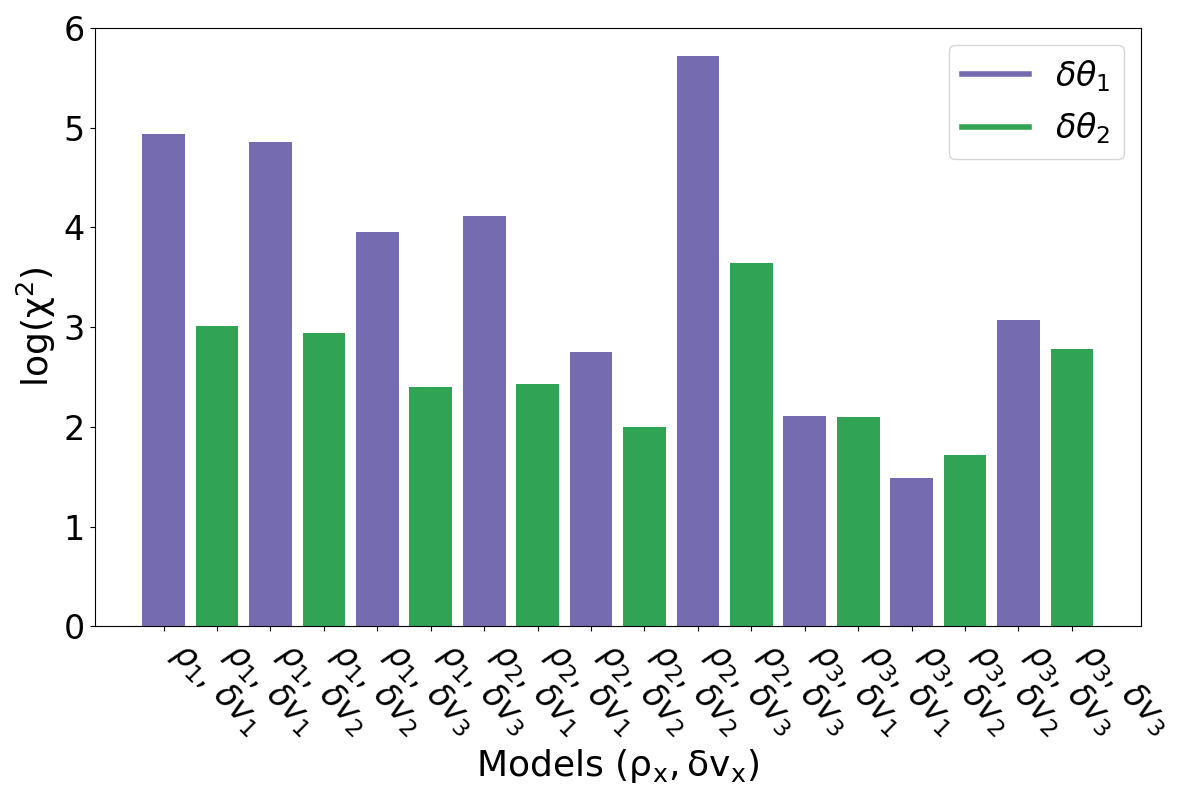}
        \caption{DCF (f=0.5)}
        \label{DCF_res}
    \end{subfigure}
    
    \vspace{0.3cm}  % Adjust vertical space as needed
    
    \begin{subfigure}{0.45\textwidth}  % Set the subfigure width (adjust as needed)
        \centering  % Center the subfigure
        \includegraphics[width=\textwidth, height=0.4\textheight, keepaspectratio]{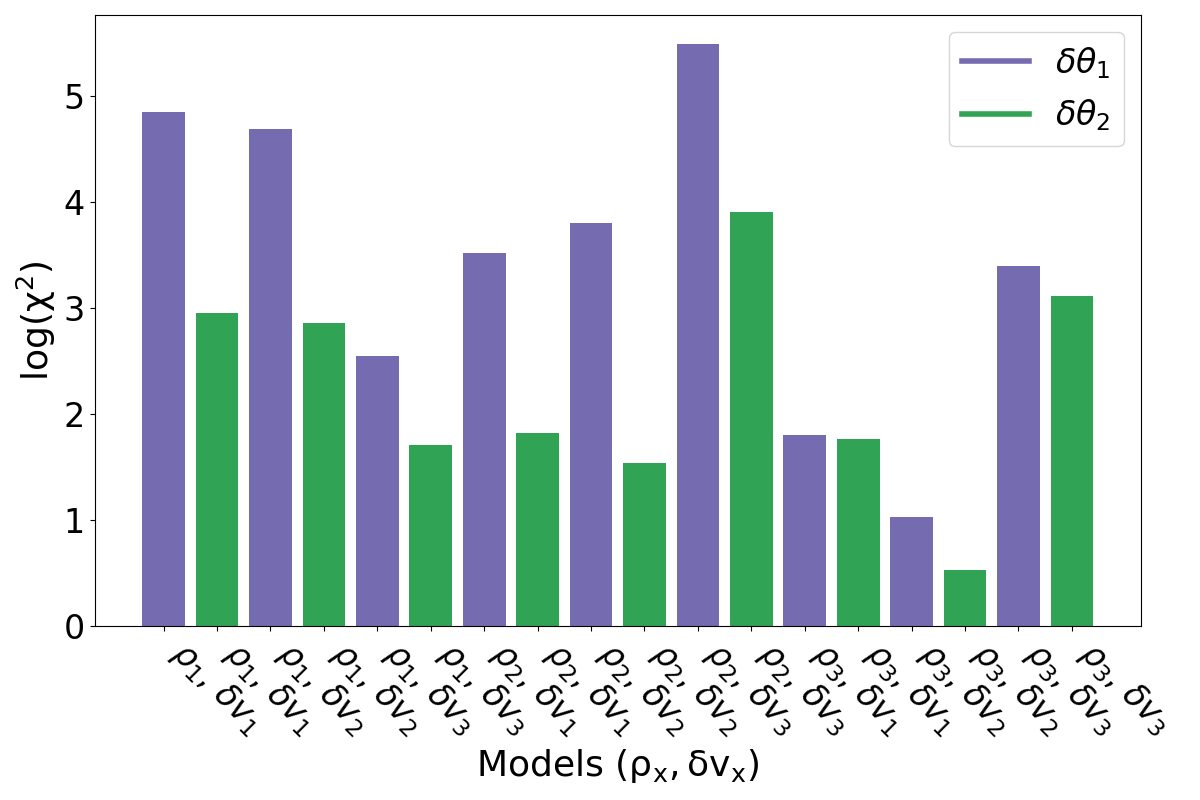}
        \caption{ST}
        \label{ST_res}
    \end{subfigure}

    \caption{The $\chi^2$ (on a logarithmic scale) for each combination of the three parameters we study. Different colors for the bars correspond to different ways that $\delta\theta$ is calculated. The indices following each parameter are explained in Table ~{\ref{table:matrix_representation}}.}
    \label{residuals}
\end{figure}

In Fig.~{\ref{residuals}} we present the $\chi^2$ for both methods and for every combination of observables $\delta$v, $\delta\theta$, $\rho$. We adjust the $y$ axis of Fig.~{\ref{residuals}} to a logarithmic scale. The labels on the horizontal axes denote the approach used to calculate $\rho$ and $\delta$v, and with the two colors we differentiate between the two approaches used
to calculate $\delta\theta$ (always according to the notation described in Table {\ref{table:matrix_representation}}). Chi-squared for some combinations can reach high values, but these are less relevant to our goal of identifying the models with the smallest deviations from the true magnetic field values.

We observe that the models that behave better, for both methods, are those where the density is estimated using RADEX, $\delta$v from the second moment maps and the difference between the methods appears to be the way that we derive the $\delta\theta$ parameter. For DCF, the lower $\chi^2$ is achieved when $\delta\theta$ is derived by taking the circular standard deviation of the polarization angle distributions, and for ST by fitting Gaussians to remove the hourglass morphology. Although we observe that the two different ways that we calculate $\delta\theta$ are close for both the methods, the lowest $\chi^2$ is achieved using the ST method. We decide to take as the best combination for both the methods, the one where every possible correction is applied: $\rho$ obtained using \textsc{RADEX} (Sect.{\ref{Dens3}}), $\delta$v derived from the second moment maps (Sect.{\ref{dv2}}), and $\delta \theta$ determined by fitting Gaussians to the polarization angle distributions (Sect.~{\ref{dtheta2}}).

To verify our results and ensure the robustness of our approach, we adjust the initial column density cutoff by increasing it by one order of magnitude to $10^{22}~\rm{cm^{-2}}$. This adjustment effectively zooms in on the cloud's central region, allowing us to assess whether the agreement with the simulation results persists and to observe any changes in the outcome. We present these results in Appendix~{\ref{appendix}}, where we see that the results lack consistency across the inclination angles, since they do not monotonically decrease. This clearly suggests that both methods lose accuracy in the innermost regions.

\section{Discussion}\label{Discussion}

In this study, we explored various approaches for calculating the three parameters ($\rho, \delta\text{v}, \delta\theta$), needed to estimate the POS component of magnetic field strength, based on the DCF and ST methods. In order to have an unbiased approach to selecting the method that performs better, we used the chi-squared metric. By comparing every possible combination, we found that ST consistently outperforms DCF. Through this process, we developed a reliable recipe that we now propose, and can be used when deriving the magnetic field strength in collapsing molecular cloud cores. This recommendation is possible because both methods achieve their best estimations when the same approach is used to calculate these parameters. If this consistency were not the case, such a recommendation would not be feasible. Specifically we encourage observers, when dealing with self-gravitating clouds, to estimate these parameters as

\begin{itemize}
    \item $\rho$: using inverse radiative-transfer analysis (Sect.~{\ref{Dens3}}),
    \item $\delta$v: from the second moment maps (Sect.~{\ref{dv2}}),
    \item $\delta\theta$: by fitting Gaussians on the polarization angles distributions (Sect.~{\ref{dtheta2}}).
\end{itemize}

We note that not all corrections to each of the observables, $\delta$v, $\delta\theta$ and $\rho$, are equally important in the final estimation of $\rm{B_{POS}}$. If we want to have an hierarchy of importance, it would be the following.

According to the $\chi^2$ values, we observe that the first and most significant impact on the magnetic field estimations is related to the estimation of the density, with better estimates achieved when it is calculated through an inverse radiative-transfer analysis. This is expected, as the two geometrical approximations discussed in Sect.~{\ref{Dens}} are over-simplistic given the lack of information regarding the cloud's true 3D geometry. On the other hand, \textsc{RADEX} relies directly on spectra to estimate the density, even though we specify the expected upper and lower limits for the column density.

The method used to estimate $\delta$v also affects the final values of $\rm{B_{POS}}$. This is evident from the considerable variation in results obtained from the different approaches tested in this work. We recommend observers to use the second moment maps to derive this parameter, as they provide less inaccurate results compared to average spectra or first moment maps, as illustrated in Fig.~{\ref{residuals}}. Calculating velocity dispersion from the average spectrum, combines all velocity components along the LOS, leading to broader distributions. In contrast, second moment maps preserve localized variations, capturing the smaller dispersions due to turbulence only that we aim to measure. Using $\delta$v from the first moment maps would significantly underestimate the values of $\rm{B_{POS}}$ and exacerbate the inconsistency with the true values for different inclination angles.

The last parameter $\delta\theta$, also has an important impact on the results. While Fig.~{\ref{residuals}} shows that varying $\delta\theta$ alone, while keeping the other parameters constant, does not drastically alter the outcomes for some cases, its role remains crucial. We can observe this effect more easily in Fig.~{\ref{fig:combined_comp}}, where, in each panel, we can directly compare the impact of this parameter for both methods. When $\delta\theta$ is calculated by fitting Gaussians to the polarization distributions, the expected cosine trend of magnetic field as a function of the inclination angle becomes evident, particularly in the ST method. This behavior is not observed when $\delta\theta$ is calculated just by considering the circular standard deviation of the polarization angle distributions, without performing any corrections. Thus this approach effectively removes the influence of the hourglass morphology.

The DCF and ST methods both probe the median of the molecular species-weighted magnetic field, rather than directly measuring the magnetic field at any specific location. This distinction is crucial because these methods provide an average estimate of the magnetic field strength over the entire cloud, which may not correspond to the conditions at the center or along a specific LOS. For example, we applied these methods to a cloud with a size of approximately 1 pc and derived one value for the magnetic field strength, but assigning this value to a density of $10^5 \ \rm{cm^{-3}}$ or a column density of $8 \times 10^{22}$ at the center of the cloud, is inaccurate. This discrepancy is evident in the distributions of the true magnetic field strength shown in Fig.~{\ref{comp_lognormal}}, where at the cloud's center, the actual magnetic field strength can be up to five times stronger than the value derived using these methods. Such differences highlight the importance of considering the averaging nature of the DCF and ST methods, especially in the context of non-uniform clouds. In Appendix \ref{appendix}, we test the best combination using a higher column density threshold and find that the results lack consistency across the inclination angles, since they do not monotonically decrease. This clearly suggests that both methods lose accuracy in the innermost regions.

\cite{VALDIVIA} assessed the use of polarized dust emission as a tracer for magnetic field topologies in low-mass protostellar cores using 27 MHD model realizations. They find that mm and submm polarized dust emission reliably traces the magnetic field in the inner protostellar envelopes, with measurements accurate to within 15$^\circ$ in 75\%-95\% of the lines of sight. Large discrepancies occur only in regions with disorganized fields or high column densities. Their results provide further evidence that using polarized dust emission can yield reliable results for tracing the magnetic field in star-forming regions.

In this study, we considered the observational uncertainties in the spectra, but we did not account for uncertainties in the Stokes parameters I, Q, and U, which could potentially impact the accuracy of the magnetic field estimation methods (e.g., \citealt{SOLER}). Accounting for these uncertainties could further refine the magnetic field strength calculations. \cite{LIU2021} used 3D MHD simulations and synthetic polarization measurements to assess the reliability of the DCF method for estimating magnetic field strengths in dense star-forming regions. They investigated the effects of beam smoothing, interferometric filtering, and other observational factors on the DCF method’s accuracy. Based on their results, we expect that if we had included them, the beam resolution of telescopes would likely lead to an underestimation of the polarization angle dispersion and thus overestimate the magnetic field strength. This overestimation is particularly noticeable when the observed region is barely resolved, meaning when the spatial scale is comparable to the beam size.

The biases in the DCF method are due to the disparity between kinetic and fluctuating magnetic energy when turbulence is compressible \citep{federrath2016, BEATTIE,BEATTIE2022,SKALIDIS} \footnote{\cite{lazarian_2025} found an excess in kinetic energy with respect to magnetic in kinematically-driven incompressible turbulence simulations. Numerical artifacts can become important with this type of driving, which artificially correlates oppositely-traveling Alfv\'en wave packets \citep{maron_goldreich_2001}, which may be responsible for the energy imbalance.}. The ST method accounts for the energy stored in the coupling between the mean and fluctuating field, which is realizable in compressible turbulence \citep{Bhattacharjee_1988, Bhattacharjee_1998}, and thus it produces more accurate estimates of the magnetic field strength. Correction factors have been employed to account for the energy inequalities in the DCF equation \citep{OSTRIKER, Li_2022, LIU2021}, but they depend on the Alfv\'en Mach number of turbulence \citep{BEATTIE2022, SKALIDIS2023}. This implies that prior knowledge of the magnetic field strength is required to accurately correct for the biases of the DCF method, rendering them of limited usefulness. On the other hand, the ST energy equipartition holds from sub- to super-Alfv\'enic turbulence \citep{BEATTIE2022, SKALIDIS2023}, and hence no extra corrections are required. This explains the consistency in the results of the ST method in the regime of ISM turbulence in both diffuse and collapsing numerical simulations (\citealt{SKALIDIS} and this work). 
 
\section{Summary}\label{Summary}

In this work, we used a chemo-dynamical simulation of a collapsing molecular cloud, when the central number density has reached a value of $10^5 \rm{cm^{-3}}$, to create synthetic observations (column densities, PPV cubes, polarization maps) under seven different inclination angles. We analyzed the synthetic observations performing corrections of increasing sophistication in the analysis of the parameters that enter the DCF and ST methods.

We determined that the magnetic field value probed by both methods corresponds to the molecular species-weighted median POS component of the true magnetic field. After examining all possible combinations of $\rho$, $\delta v$, and $\delta \theta$, we found that the method developed by \cite{ST} achieves higher accuracy, as it more closely follows the expected cosine trend with respect to the inclination angle, and consistently remains within 1$\sigma$ from the median of the molecular species-weighted POS component of the magnetic field.

To identify the optimal combination of corrections for $\rho$, $\delta v$, and $\delta\theta$, we employed the $\chi^2$ metric, aiming to quantify the difference from the ``true'' magnetic field strength. The best combination we identified is as follows: $\rho$ obtained using inverse radiative-transfer analysis (Sect.{\ref{Dens3}}), $\delta$v derived from the second moment maps (Sect.{\ref{dv2}}), and $\delta \theta$ determined by fitting Gaussians to the polarization angle distributions (Sect.~{\ref{dtheta2}}). We suggest that observers adopt this methodology when studying self-gravitating clouds.

\begin{acknowledgements}

We thank the anonymous referee for comments which significantly helped to improve this work. A. Polychronakis and K. Tassis acknowledge support from the European Research Council (ERC) under the European Unions Horizon 2020 research and innovation programme under grant agreement No. 7712821. A. Tritsis acknowledges support by the Ambizione grant no. PZ00P2\_202199 of the Swiss National Science Foundation (SNSF). Support for this work was provided by NASA through the NASA Hubble Fellowship grant \#HST-HF2-51566.001 awarded by the Space Telescope Science Institute, which is operated by the Association of Universities for Research in Astronomy, Inc., for NASA, under contract NAS5-26555. The software used in this work was in part developed by the DOE NNSA-ASC OASCR Flash Center at the University of Chicago. We also acknowledge use of the following software: \textsc{Matplotlib} (\citealt{HUNTER}), \textsc{Numpy} (\citealt{HARRIS}), \textsc{Scipy} (\citealt{VIRTANEN}), \textsc{Numba} (\citealt{LAM}), and the \textsc{yt} analysis toolkit (\citealt{TURK}).

\end{acknowledgements}


\bibliographystyle{aa}  % A&A journal style for the bibliography
% \bibliography{bibliography}  

\clearpage

\appendix
\section{Results for the higher column density threshold}
\label{appendix}

\begin{figure}[t]
    \centering
    \includegraphics[width=\textwidth, height=0.6\textheight, keepaspectratio]{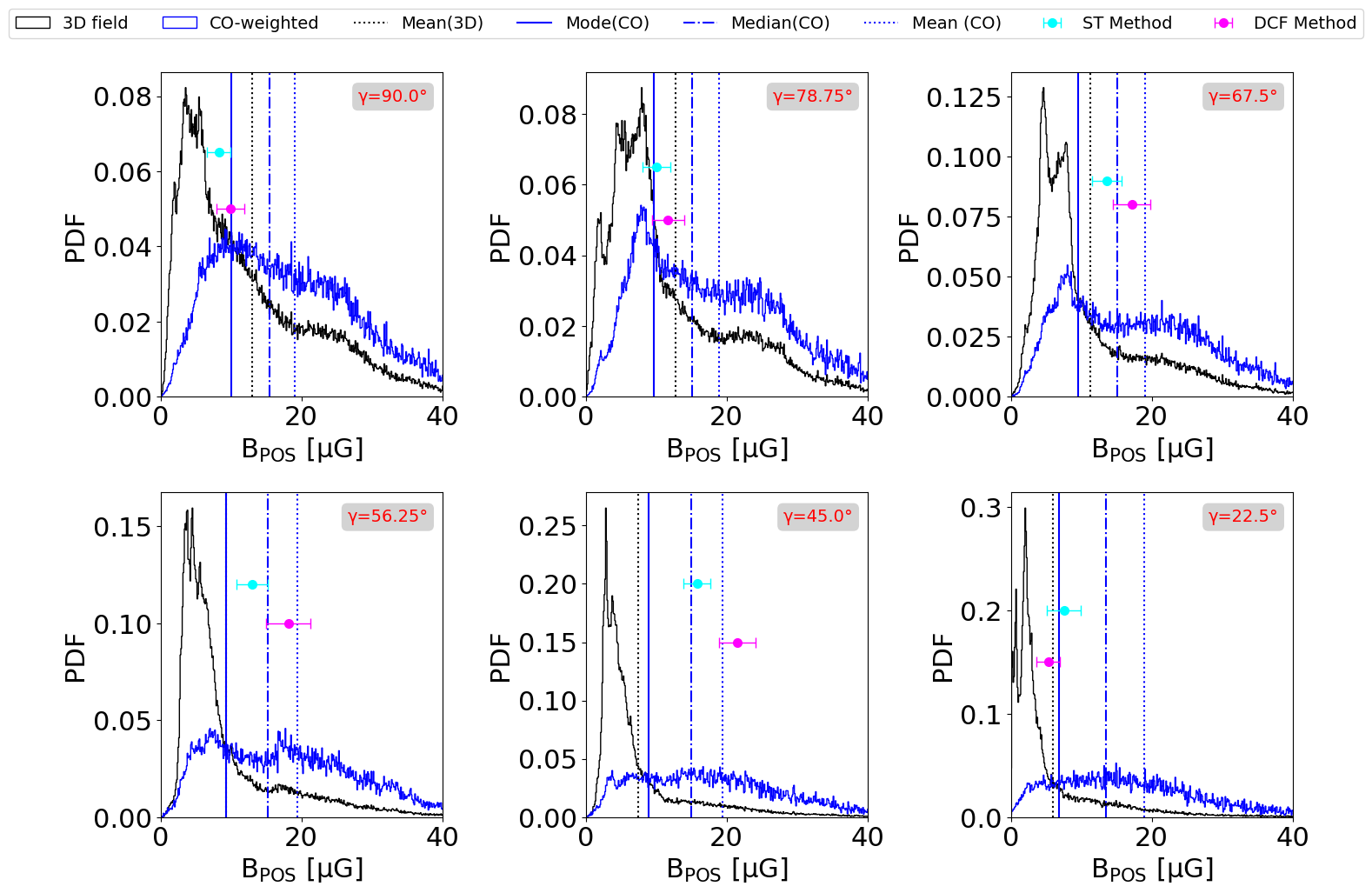}
    \vspace{0cm} % Adjust vertical position
    \hspace{0cm} % Adjust horizontal position
    \parbox{\textwidth}{ % Adjust the caption box width
        \captionsetup{width=1.0\textwidth, justification=justified, singlelinecheck=false} % Caption width
        \caption{Distributions of $\rm{B_{POS}}$ from the simulation for different inclination angles $\gamma$. For each inclination angle we have only considered the locations where the column density exceeds the updated threshold ($10^{22} \rm{cm^{-2}}$). Black distributions show the 3D values of the $\rm{B_{POS}}$. Blue distributions show the CO-weighted values of $\rm{B_{POS}}$. Solid, dash-dotted and dotted lines mark the mode, median and mean of each distribution respectively. Points show the values that the methods we are testing predicted.}
        \label{comp_lognormal_new}
    }
\end{figure}

In Fig.~{\ref{comp_lognormal_new}}, we present the results of the simulations when a threshold at $10^{22} \,\mathrm{cm^{-2}}$ is set in column density, therefore all synthetic data that fall in regions of the cloud that do not exceed this column density threshold, are removed from our analysis. The points, similarly to Fig.~{\ref{comp_lognormal}}, refer to the methods' predictions. This time we show just one combination which is: $\rho$ using inverse radiative-transfer analysis, $\delta$v from the second moment maps, and $\delta\theta$ by fitting Gaussians to the polarization angles distributions. We note here that for inclination angles $0^\circ$ and $22.5^\circ$ we only fit one, instead of two Gaussians, given that the shape of the distributions for this angles does not allow for the inclusion of a second Gaussian component.

The derived values are within the ranges of the simulation but as opposed to the previous column density threshold, there is no consistency across the inclination angles, since they do not monotonically decrease, which is a clear indication that both methods lose accuracy in inner (dense) regions.

\clearpage
\appendix
\setcounter{section}{1} % This sets Appendix B (A=0, B=1)
\renewcommand{\thesection}{\Alph{section}} % Format appendix as "Appendix A", "Appendix B", etc.
\renewcommand{\thefigure}{\Alph{section}.\arabic{figure}} % Format figures as A.1, B.1, etc.
\setcounter{figure}{0}

% Begin actual appendix section
\section{Procedure to derive $\delta\theta$ using a dispersion function}
\label{appendix2}

% If needed, add this to include the section in the TOC
\addcontentsline{toc}{section}{Appendix \Alph{section}: Structure function - based derivation of $\delta\theta$}

\begin{figure}[t]
    \centering
    \includegraphics[width=\textwidth, height=0.6\textheight, keepaspectratio]{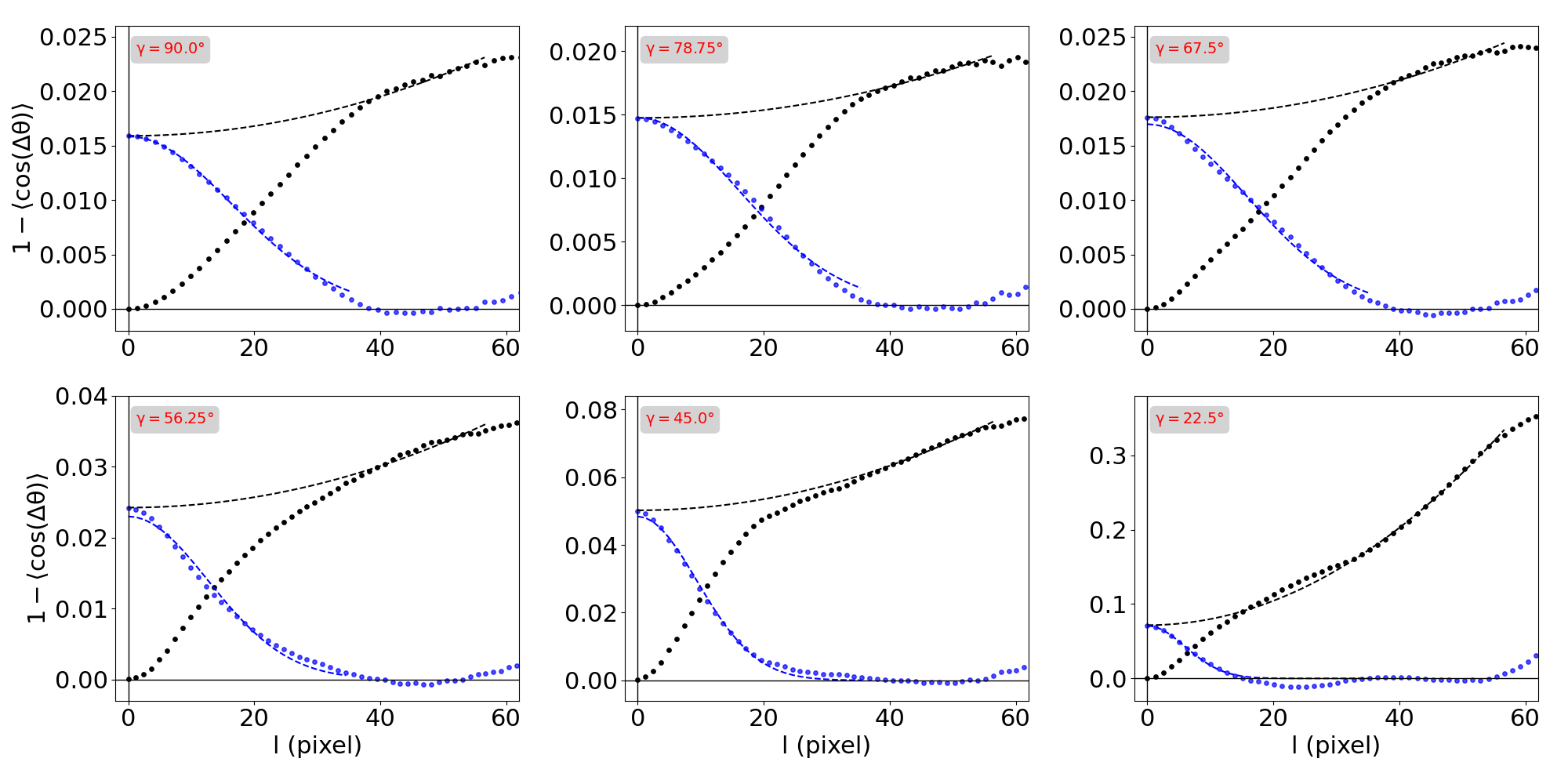}
    \vspace{0cm} % Adjust vertical position
    \hspace{0cm} % Adjust horizontal position
    \parbox{\textwidth}{ % Adjust the caption box width
        \captionsetup{width=1.0\textwidth, justification=justified, singlelinecheck=false} % Caption width
        \caption{Dispersion functions for various inclination angles. In each panel, the black dots correspond to the dispersion function computed using Eq. \ref{eqB1} on every panel. Different panels refer to different inclination angles. The model fit at large scales is shown with the black dashed curve. Blue points mark the dispersion function with the black line subtracted, and the blue line shows the fit of Eq. \ref{eqB2} (only the exponential term).}
        \label{Houde_fits}
    }
\end{figure}

According to the procedure analytically explained in \cite{HOUDE}, an alternative way for estimating the $\rm{\delta B/B}$ ratio, is the following. First we calculate the dispersion function for each polarization map, as

\begin{equation}\label{eqB1}
    \rm{\langle \cos[\Delta \phi(l)] \rangle = \langle \cos[\Phi(x) - \Phi(x + l)] \rangle},
\end{equation}
where $\rm{\Phi(x)}$ denotes the polarization angle measured in radians, x denotes the 2D coordinates in the POS, l the spatial separation of two polarization measurements in the POS, and brackets denote averaging over the entire polarization map. The polarization angle differences are constrained in  the interval $[0, 90^\circ]$.

HH09 defined the total magnetic field as \( \rm{B_{\text{tot}} = B_0 + B_t} \), where \( \rm{B_0} \) represents the mean magnetic field component and \( \rm{B_t} \) is the turbulent (or random) component. They assumed that the strength of \( \rm{B_0} \) is uniform, while \( \rm{B_t} \) is generated by turbulent gas motions. After these assumptions, they derived the following analytical relation for Eq. \ref{eqB1},

\begin{align}\label{eqB2}
\rm{1 - \langle \cos[\Delta \phi(l)] \rangle} &\simeq 
\rm{\sqrt{2\pi} \frac{\langle B_t^2 \rangle}{B_0^2} 
\frac{\delta^3}{(\delta^2 + 2W^2)\Delta'} 
\left(1 - e^{-l^2 / 2(\delta^2 + 2W^2)}\right)} \notag \\
&\quad + \rm{m l^2},
\end{align}

\needspace{6\baselineskip}
\vspace*{32em}

\noindent where \( \rm{B_0} \) has its usual meaning as the mean magnetic field component, while \( \rm{B_t} \) refers to the turbulent component of the magnetic field. W denotes the beam size, \( \rm{\Delta'} \) represents the effective cloud depth, \( \rm{\delta} \) is the correlation length, and m is a constant that is determined by fitting the dispersion function to the data. Consequently, the term $\rm{\langle B_t^2 \rangle / B_0^2}$ is equivalent to \( \delta\theta^2 \), which is the quantity we aim to constrain. In our synthetic observations, we assume a pencil beam, meaning that W is set to zero. Regarding \( \rm{\Delta'} \), HH09 note that it is less than or equal to the depth of the cloud. In this work, we adopt a value for \( \rm{\Delta'} \) equal to the POS dimension of the cloud for each inclination angle. However, it is important to emphasize that this is somewhat arbitrary, as determining the effective cloud depth involves significant uncertainties.

To apply the method, we first calculate the dispersion function using Eq. \ref{eqB1}, and then we fit the model given on the right-hand a side of Eq. \ref{eqB2}, which includes three free parameters: $\rm{\langle B_t^2 \rangle / B_0^2}$, $\delta$, and m. Using our synthetic data, we computed the dispersion function, shown as black dots in Fig. \ref{Houde_fits}. We then fit the right-hand side of Eq. \ref{eqB2} to these data points, omitting the exponential term from the model. This fit, indicated by the black dashed curve, was performed at larger spatial scales, i.e., for $\rm{35 < l < 55}$  pixels (approximately from 0.5 to 0.9 pc). It is important to note that this range remained consistent for all the different inclination angles $\gamma$. The fitted curve was subsequently subtracted from the original data, and the residuals—shown as blue dots—represent the turbulent auto-correlation function, corresponding to the second term of Eq. \ref{eqB2}. We then fit the previously omitted term (only the exponential term) to the residuals (blue points), allowing us to derive the desired value of $\delta\theta$. We repeat the same process for every different inclination angle $\gamma$. The results of the fits are also shown in Fig. \ref{Houde_fits}.

\end{document}